\pdfoutput=1

\documentclass[11pt,twoside,a4paper,cmspaper,final,collab]{cms-tdr}
\def\svnVersion{376960}\def\cmsCernNoTag{CERN-EP/2016-112}\def\cmsCernDate{\today}\def\cmsMessage{Published in Physics Letters B as \href{http://dx.doi.org/10.1016/j.physletb.2016.09.062}{\doi{10.1016/j.physletb.2016.09.062.}}}

\begin{document}\cmsNoteHeader{HIG-14-040}

\hyphenation{had-ron-i-za-tion}
\hyphenation{cal-or-i-me-ter}
\hyphenation{de-vices}

\RCS$Revision: 376949 $
\RCS$HeadURL: svn+ssh://svn.cern.ch/reps/tdr2/papers/HIG-14-040/trunk/HIG-14-040.tex $
\RCS$Id: HIG-14-040.tex 376949 2016-12-12 18:22:50Z cjessop $

\ifthenelse{\boolean{cms@external}}{\providecommand{\cmsLLeft}{Top\xspace}}{\providecommand{\cmsLLeft}{Top left\xspace}}
\ifthenelse{\boolean{cms@external}}{\providecommand{\cmsCenter}{Middle\xspace}}{\providecommand{\cmsCenter}{Top right\xspace}}
\ifthenelse{\boolean{cms@external}}{\providecommand{\cmsRRight}{Bottom\xspace}}{\providecommand{\cmsRRight}{Bottom\xspace}}
\ifthenelse{\boolean{cms@external}}{\providecommand{\cmsLeft}{top\xspace}}{\providecommand{\cmsLeft}{left\xspace}}
\ifthenelse{\boolean{cms@external}}{\providecommand{\cmsRight}{bottom\xspace}}{\providecommand{\cmsRight}{right\xspace}}
\providecommand{\cmsTable}[1]{\resizebox{\columnwidth}{!}{#1}}
\providecommand{\CLs}{\ensuremath{\mathrm{CL}_\mathrm{s}}\xspace}
\newcommand{\hemu}{\ensuremath{\PH\to\Pe\Pgm}\xspace}
\providecommand{\mt}{\ensuremath{M_\mathrm{T}}\xspace}
\providecommand{\tauh}{\ensuremath{\Pgt_\mathrm{h}}\xspace}
\providecommand{\Hteth}{\ensuremath{\PH \to \Pgt_{\Pe} \tauh}\xspace}
\cmsNoteHeader{HIG-14-005 v1.0}
\title{Search for lepton flavour violating  decays of the Higgs boson to  $\Pe\tau$ and  $\Pe\mu$ in
proton-proton collisions at $\sqrt{s}=8$\TeV}

\date{\today}

\abstract{A direct search for lepton flavour violating decays of the Higgs boson (H) in the $\PH \to \Pe \tau$ and $\PH \to \Pe \mu$ channels is described. The data sample used in the search was collected in proton-proton collisions at $\sqrt{s}=8$\TeV with the CMS detector  at the LHC and corresponds to an integrated luminosity of 19.7\fbinv.
No evidence is found for lepton flavour violating decays in either final state. Upper limits on the branching fractions, $\mathcal{B}(\PH \to \Pe \tau )<0.69\%$ and $\mathcal{B}(\PH \to \Pe \mu)< 0.035\%$, are set at the 95\% confidence level. The constraint set on
$\mathcal{B}(\PH \to \Pe \tau)$ is an order of magnitude more stringent than the existing indirect limits. The limits are used to constrain the corresponding flavour violating  Yukawa couplings, absent in the standard model.}

\hypersetup{%
pdfauthor={CMS Collaboration},%
pdftitle={Search for lepton flavour violating  decays of the Higgs boson to  e tau and  e mu in
proton-proton collisions at sqrt(s)=8 TeV},%
pdfsubject={CMS},%
pdfkeywords={CMS, physics, Higgs, electrons, lepton-flavour-violation}}

\maketitle
\section{Introduction}
The discovery of the Higgs boson~\cite{Aad:2012tfa, Chatrchyan:2012ufa, Chatrchyan:2013lba} has generated great  interest in exploring
its properties. In the standard model (SM), lepton flavour violating (LFV)  decays of the Higgs boson are
forbidden.  Such decays can  occur naturally in models with more than one Higgs boson
doublet~\cite{PhysRevLett.38.622}. They also arise in supersymmetric models~\cite{DiazCruz:1999xe,Han:2000jz,Arganda:2004bz,Arhrib:2012ax,Arana-Catania:2013xma,Arganda:2015uca,Arganda:2015naa}, composite Higgs  models~\cite{Agashe:2009di,Azatov:2009na}, models with flavour
symmetries~\cite{Ishimori:2010au}, Randall--Sundrum models~\cite{Perez:2008ee,Casagrande:2008hr,Buras:2009ka}, and
others ~\cite{Blanke:2008zb,Giudice:2008uua,AguilarSaavedra:2009mx,Albrecht:2009xr,Goudelis:2011un,McKeen:2012av,
Pilaftsis199268,PhysRevD.47.1080,Arganda:2014dta}.  The CMS Collaboration has recently published a search in the $\PH \to \Pgm \Pgt$
channel~\cite{Khachatryan:2015kon} showing an excess of data with respect to the SM background-only hypothesis at
$M_{\PH} =125\GeV$ with a significance of $2.4$ standard deviations ($\sigma$). A constraint is set on the branching fraction  $\mathcal{B}(\PH \to \Pgm \Pgt)<1.51\%$
at 95\% confidence level (CL), while the best fit branching fraction is $\mathcal{B}(\PH \to \Pgm \Pgt)=(0.84^{+0.39}_{-0.37})\%$. The ATLAS Collaboration finds a deviation from the background expectation of $1.3\sigma$ significance
in the $\PH \to \Pgm \Pgt$ channel  and sets an upper limit of  $\mathcal{B}(\PH \to \Pgm \Pgt)<1.85\%$ at 95\% CL
with a best fit branching fraction of $\mathcal{B}(\PH \to \Pgm \Pgt)=(0.77 \pm 0.62)\%$~\cite{Aad:2015gha}. To date, no dedicated searches have been published for the $\PH \to \Pe \Pgm$  channel. The ATLAS collaboration recently reported  searches for $\PH \to \Pe \Pgt$ and $\PH \to \Pgm \Pgt$, finding no significant excess of events over the background expectation. The searches in channels with leptonic tau decays are sensitive only to a difference between $\mathcal{B}(\PH \to \Pe \Pgt)$ and $\mathcal{B}(\PH \to \Pgm \Pgt)$. These are combined with the searches in channels with hadronic tau decays to set limits of $\mathcal{B}(\PH \to \Pe \Pgt)<1.04\%$, $\mathcal{B}(\PH \to \Pgm \Pgt)<1.43\%$ at 95\% CL~\cite{Aad:2016blu}. There are also indirect constraints. The presence of  LFV Higgs boson couplings allows,
$\Pgm \to \Pe$, $\Pgt \to \Pgm$, and $\Pgt \to \Pe$ to proceed via a virtual Higgs
boson~\cite{McWilliams:1980kj,Shanker:1981mj}. The experimental limits  on these decays have  been
translated into constraints on $\mathcal{B}(\PH \to \Pe \Pgm)$, $\mathcal{B}(\PH \to \Pgm \Pgt)$ and
$\mathcal{B}(\PH \to \Pe \Pgt)$~\cite{Blankenburg:2012ex,Harnik:2012pb}.  The null result for
$\Pgm \to \Pe \gamma$~\cite{Beringer:1900zz} strongly constrains $\mathcal{B}(\hemu) < \mathcal{O}(10^{-8})$. However, the constraint $\mathcal{B}(\PH \to \Pe \Pgt) < \mathcal{O}(10\%)$ is much less stringent. This comes from  searches for rare $\tau$ decays~\cite{Celis:2013xja}
such as $\Pgt \to \Pe \gamma$, and the measurement of the electron magnetic moment.  Exclusion limits on the
electric dipole moment of the electron~\cite{Barr:1990vd} also provide complementary constraints.

This letter describes a search  for LFV decays of the Higgs boson with $M_{\PH}=125$\GeV, based
on proton-proton collision data recorded at $\sqrt{s}=8\TeV$ with the CMS detector at the CERN LHC, corresponding to an integrated
luminosity of 19.7\fbinv. The search is performed in  three  decay channels,  $\PH \to \Pe \Pgt_{\Pgm}$,
$\PH \to \Pe \tauh$, and $\PH \to \Pe \Pgm$, where $\Pgt_{\Pgm}$ and $\tauh$ correspond to muonic and hadronic decay channels of tau leptons, respectively.  The decay channel, $\PH \to \Pe \Pgt_{\Pe}$, is
not considered due to the large background contribution from $\cPZ \to \Pe \Pe$ decays. The expected final state signatures are  very similar to the
SM $\PH \to {\Pgt_{\Pe}} \tauh$ and $\PH \to {\Pgt_{\Pe}} \Pgt_{\Pgm}$
decays, studied by CMS~\cite{Chatrchyan:2014vua,CMS-PAPERS-HIG-13-004} and ATLAS~\cite{Aad:2015vsa},
but with some significant kinematic differences. The electron in the LFV $\PH \to \Pe \Pgt$ decay
is produced promptly, and tends to have a larger momentum than in the SM $\PH \to \Pgt_{\Pe}\tauh$ decay. In the \hemu channel, $M_{\PH}$ can be measured with good resolution due to the absence of neutrinos.

This letter is organized as follows. After a description of the CMS detector (Section \ref{cmsdet}) and of the collision data and simulated samples used in the analysis (Section \ref{samples}), the event reconstruction is described in Section~\ref{reconst}. The event selection and the estimation of the background and its components are described separately for the two Higgs decay modes $\PH \to \Pe \Pgt$ and $\PH \to \Pe \Pgm$ in Sections~\ref{etau} and \ref{emu}. The results are then presented in Section~\ref{results}.

\section{The CMS detector\label{cmsdet}}

A detailed description of the CMS detector, together with a definition of the coordinate system used and the relevant kinematic variables, can be found in Ref.~\cite{CMS-JINST}. The momenta of charged particles are
measured with a silicon pixel and strip tracker that covers the
pseudorapidity range $\abs{\eta} < 2.5$, in a 3.8\unit{T}
axial magnetic field.
A lead tungstate crystal electromagnetic calorimeter
(ECAL) and a brass and scintillator hadron calorimeter,
both consisting of a barrel section and two endcaps, cover the pseudorapidity range  $\abs{\eta} < 3.0$. A
steel and quartz-fibre Cherenkov forward detector extends the calorimetric
coverage to $\abs{\eta} < 5.0$. The outermost component of the CMS detector is the
muon system, consisting of gas-ionization detectors placed in the
steel flux-return yoke of the magnet to identify the muons traversing the detector. The two-level CMS trigger system selects events of interest for
permanent storage. The first trigger level,
composed of custom hardware processors, uses information from the
calorimeters and muon detectors to select events in less than 3.2\mus.
The software algorithms of the high-level trigger, executed on a farm of
commercial processors, reduce the event rate to less than 1\unit{kHz} using information from all detector subsystems.

\section{Collision data and simulated events \label{samples}}
The triggers for the $\PH \to \Pe \Pgt_{\Pgm}$  and  $\hemu$ analyses require an electron and a muon candidate.
The trigger for $\PH \to \Pe \tauh$ requires a single electron. More details on the trigger selection are given in Sections~\ref{etauEvtSel} and \ref{emuEvtSel}, for the $\PH \to \Pe \Pgt$  and  $\hemu$ channels  respectively.
Simulated samples of signal and background events are produced with several event generators. The CMS detector response is modelled using \GEANTfour~\cite{GEANT4}. The Higgs bosons are produced in proton-proton collisions predominantly by gluon fusion (GF)~\cite{Georgi:1977gs}, but also by vector boson fusion (VBF)~\cite{Cahn:1986zv} and in  association with a $\PW$ or $\cPZ$ boson~\cite{Glashow:1978ab}.
The $\PH \to \Pe \Pgt$  decay sample is produced with \PYTHIA 8.176~\cite{Sjostrand:pythia8} using
the CTEQ6L parton distribution functions (PDF). The $\PH \to \Pe \Pgm$ decay sample is  produced with
\PYTHIA 6.426~\cite{pythia} using the CT10 parton distribution functions~\cite{Nadolsky:2008zw}.
The SM Higgs boson samples are generated using \POWHEG 1.0~\cite{Nason:2004rx,Frixione:2007vw, Alioli:2010xd, Alioli:2010xa, Alioli:2008tz}, with CT10 parton distribution functions, interfaced to \PYTHIA 6.426.
The \MADGRAPH 5.1.3.30~\cite{Alwall:2011uj}
generator is used for $\cPZ\mathrm{+jets}$, $\PW\text{+jets}$, top anti-top quark pair production $\ttbar$, and diboson production, and \POWHEG for single top quark production. The \POWHEG and \MADGRAPH generators are interfaced to \PYTHIA 6.426 for
parton shower and hadronization.
The \PYTHIA parameters for the underlying event description are set to the Z2* tune.
The Z2* tune is derived from the Z1 tune~\cite{Field:2010bc}, which uses the CTEQ5L parton distribution set,
whereas Z2* adopts CTEQ6L.
Due to the high luminosities attained during data-taking, many events have multiple proton-proton interactions per bunch crossing (pileup). All simulated samples are reweighted to match the pileup distribution observed in
data.

\section{Event reconstruction \label{reconst}}
Data were collected at an  average pileup of 21 interactions per bunch crossing.
The tracking system is able to separate collision vertices as close as 0.5\mm to each other along the beam  direction~\cite{Chatrchyan:2014fea}.
The primary vertex, assumed to correspond to the hard-scattering process, is the vertex for which the sum of the squared transverse momentum $\pt^2$ of all the associated tracks is the largest. The pileup interactions  also affect the identification of most of the physics objects, such as jets, and variables such as lepton isolation.

A particle-flow (PF)  algorithm~\cite{CMS-PAS-PFT-09-001, CMS-PAS-PFT-10-002, CMS-PAS-PFT-10-003} combines  the information from all CMS subdetectors to
identify and reconstruct the individual particles emerging from all
interactions in the event: charged and neutral hadrons, photons, muons, and electrons.
These particles are then required to be consistent with the primary vertex
and used to reconstruct jets, hadronic $\Pgt$ decays, quantify the isolation of leptons and
photons and reconstruct \ETmiss. The missing transverse energy vector, $\VEtmiss$, is defined as the negative of the vector sum of the \pt of all identified PF objects in the event~\cite{Khachatryan:2014gga}. Its magnitude is referred to as \ETmiss.
 The variable $\Delta R = \sqrt {\smash[b]{(\Delta\eta)^2 +(\Delta\phi)^2}}$, where $\phi$ is the azimuthal co-ordinate, is used to
measure the separation between reconstructed objects in the detector.

Electron reconstruction requires the matching of an energy cluster in the ECAL with a track in the silicon
tracker~\cite{Khachatryan:2015hwa}.
Electron candidates are accepted in the range $\abs{\eta}<2.5$, with the exception
of the region $1.44 < \abs{\eta} < 1.56$ where service infrastructure for the detector
is located.
Electron identification uses a multivariate discriminant  that combines observables sensitive to the
amount of bremsstrahlung along the electron trajectory, the
geometrical and momentum matching between the electron trajectory and
associated clusters, and shower-shape observables. Additional requirements are imposed to remove electrons produced by photon conversions. The electron energy is corrected for imperfection of the reconstruction using a regression based on a boosted decision tree~\cite{HIG-13-002}.

Muon candidates are obtained from combined fits of tracks in the tracker and muon detector seeded by track segments in the muon detector alone,
including compatibility with small energy depositions in the calorimeters.
Identification is based on track quality and  isolation.
The muon momentum is estimated with the combined fit. Any
possible bias in the measured muon momentum is determined from the position of the $\cPZ \to \Pgm\Pgm$
 mass peak as a function of muon kinematic variables, and a small correction is obtained using the procedure
described in Ref.~\cite{ref:Muscle}.

Hadronically decaying $\Pgt$ leptons are reconstructed and
identified using an algorithm~\cite{Khachatryan:2015dfa}
that selects the  decay modes with one charged hadron and up to two neutral pions,
or three charged hadrons. A photon from a neutral-pion decay can convert in the tracker
material into an electron-positron pair, which can then radiate
photons. These particles give rise to several ECAL energy
deposits at the same $\eta$ value but separated in $\phi$.
They  are reconstructed as several photons by the PF algorithm. To increase
the acceptance for these converted photons, the neutral pions are
identified by clustering the reconstructed photons in narrow strips along
the $\phi$ direction.
The charge of $\tauh$ candidates is reconstructed by summing the charges of all particles included in the construction of the
candidate, except for the electrons contained in strips.
Dedicated discriminators  veto against electrons and muons.

Jets misidentified as electrons, muons or taus are suppressed by imposing  isolation requirements, summing the
neutral and charged particle contributions in cones of $\Delta R$ about the lepton.
The energy deposited within the isolation cone is contaminated by energy from pileup and
the underlying event.
The effect of pileup is
reduced by requiring the tracks considered in the isolation sum to be compatible with originating from the production vertex of the lepton. The contribution to the isolation from pileup and the underlying event is subtracted on an event-by-event
basis. In the case of electrons, this contribution is estimated from the product of the measured energy density $\rho$ for the event, determined using the $\rho$
median estimator implemented in \FASTJET~\cite{Cacciari:2011ma}, and an effective area
corresponding to the isolation cone. In the case of muons and hadronically decaying $\Pgt$ leptons,
it is estimated on a statistical basis through the modified $\Delta \beta$
correction described in Ref.~\cite{Khachatryan:2015dfa}.

Jets  are reconstructed from all the particles using the anti-\kt jet clustering
algorithm~\cite{Cacciari:2008gp} implemented in \FASTJET, with a distance parameter of $\Delta R = 0.5$.
The jet energies are corrected by subtracting the contribution of particles
created in pileup interactions and in the underlying event~\cite{CMS-JME-10-011}.
Particles from different pileup vertices can be clustered into a pileup jet, or significantly overlap a
jet from the primary vertex below the selected jet \pt threshold. These jets are identified and
removed~\cite{CMS-PAS-JME-13-005}.

\section{\texorpdfstring{$\PH \to \Pe \Pgt$ analysis}{H -> e tau analysis} \label{etau}}
\subsection{Event selection \label{etauEvtSel}}
The $\PH \to \Pe \tauh$ selection begins by  requiring an event recorded with a single electron
trigger ($\pt^{\Pe}>27$\GeV, $\abs{\eta^{\Pe}} < 2.5$). The $\PH \to \Pe \Pgt_{\Pgm}$ channel
requires a muon-electron trigger ($\pt^{\Pe}>17$\GeV, $\abs{\eta^{\Pe}} < 2.5$, $\pt^{\Pgm}>8$\GeV,
$\abs{\eta^{\Pgm}} < 2.4$). The triggers also apply loose identification and isolation requirements to the leptons.

A loose selection is then made for both channels. Electron, muon and hadronic tau lepton candidates
are required to be isolated and to lie in the pseudorapidity ranges where they can be well reconstructed; $\abs{\eta^{\Pe}} <1.44$ or $1.57<\abs{\eta^{\Pe}} <2.30$, $\abs{\eta^{\Pgm}} <2.1$ and $\abs{\eta^{\tauh}} <2.3$, respectively. Leptons are also required to be compatible with the primary vertex and to be separated by $\Delta R > 0.4$ from any jet in the event with $\pt >$ 30\GeV. The  $\PH \to \Pe \Pgt_{\Pgm}$  channel then requires an
electron ($\pt^{\Pe} > 40$\GeV) and an oppositely charged muon ($\pt^{\Pgm} > 10$\GeV) separated by
$\Delta R >0.1$. Events in this channel  with additional muons ($\pt>7\GeV$) or
electrons ($\pt>7\GeV$) are also rejected. The  $\PH \to \Pe \tauh$ channel requires an
electron ($\pt^{\Pe} > 30$\GeV) and an  oppositely charged  hadronic tau lepton  ($\pt^{\tauh} > 30$\GeV).
Events in this channel with additional muons ($\pt>5\GeV$), electrons ($\pt>10\GeV$), or hadronic tau
leptons ($\pt>20\GeV$)  are rejected.

The events are then divided into categories within each channel according to the number of
jets in the event. Jets are required to pass identification criteria, have $\pt> 30$\GeV, and
lie in the region $\abs{\eta} < 4.7$. The 0-jet and 1-jet categories contain events primarily produced by GF.
The 2-jet category is defined to enrich the contribution from events produced via the VBF process.

The main observable used to discriminate between the signal and the background  is the collinear
mass, $M_{\text{col}}$, which provides an estimate of $M_{\PH}$ using the observed
decay products. It is constructed using the collinear approximation based on
the observation that, since \mbox{$M_{\PH}\gg M_{\Pgt}$}, the $\Pgt$ decay products are
highly Lorentz boosted in the direction of the  $\Pgt$~\cite{Ellis:1987xu}.
The neutrino momenta
can be approximated to have the same
direction as the other visible decay products of the $\Pgt$ ($\Pgt^\text{vis}$)
and the component of the $\VEtmiss$ in the direction of the visible $\tau$ decay products is used to estimate the transverse component of the neutrino momentum ($\pt^{\nu,~\text{est}}$).
The collinear mass can then be derived from the visible mass of the $\Pgt$-$\Pe$ system ($M_{\text{vis}}$) as $M_{\text{col}}= M_{\text{vis}} / \sqrt{x_{\Pgt}^\text{vis}}$, where $x_{\Pgt}^\text{vis}$ is the fraction of energy carried by the visible decay products of the $\Pgt$ ($x_{\Pgt}^\text{vis}={\pt^{\Pgt^{\text{vis}}}}/{(\pt^{\Pgt^\text{vis}}+\pt^{\nu,~\text{est}})}$).

\begin{figure*}[!hbtp]\centering
 \includegraphics[width=0.4565\textwidth]{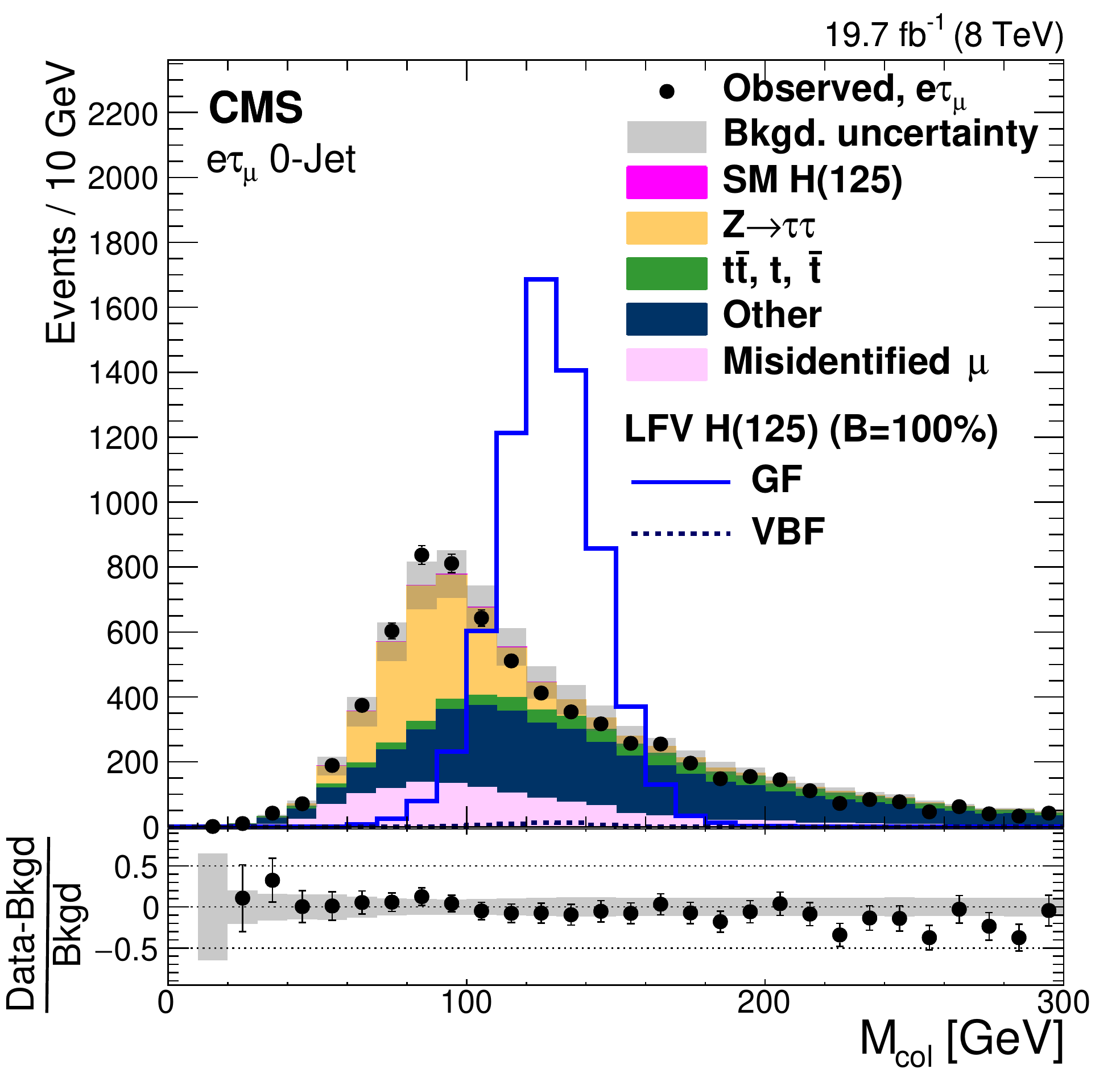}
 \includegraphics[width=0.4565\textwidth]{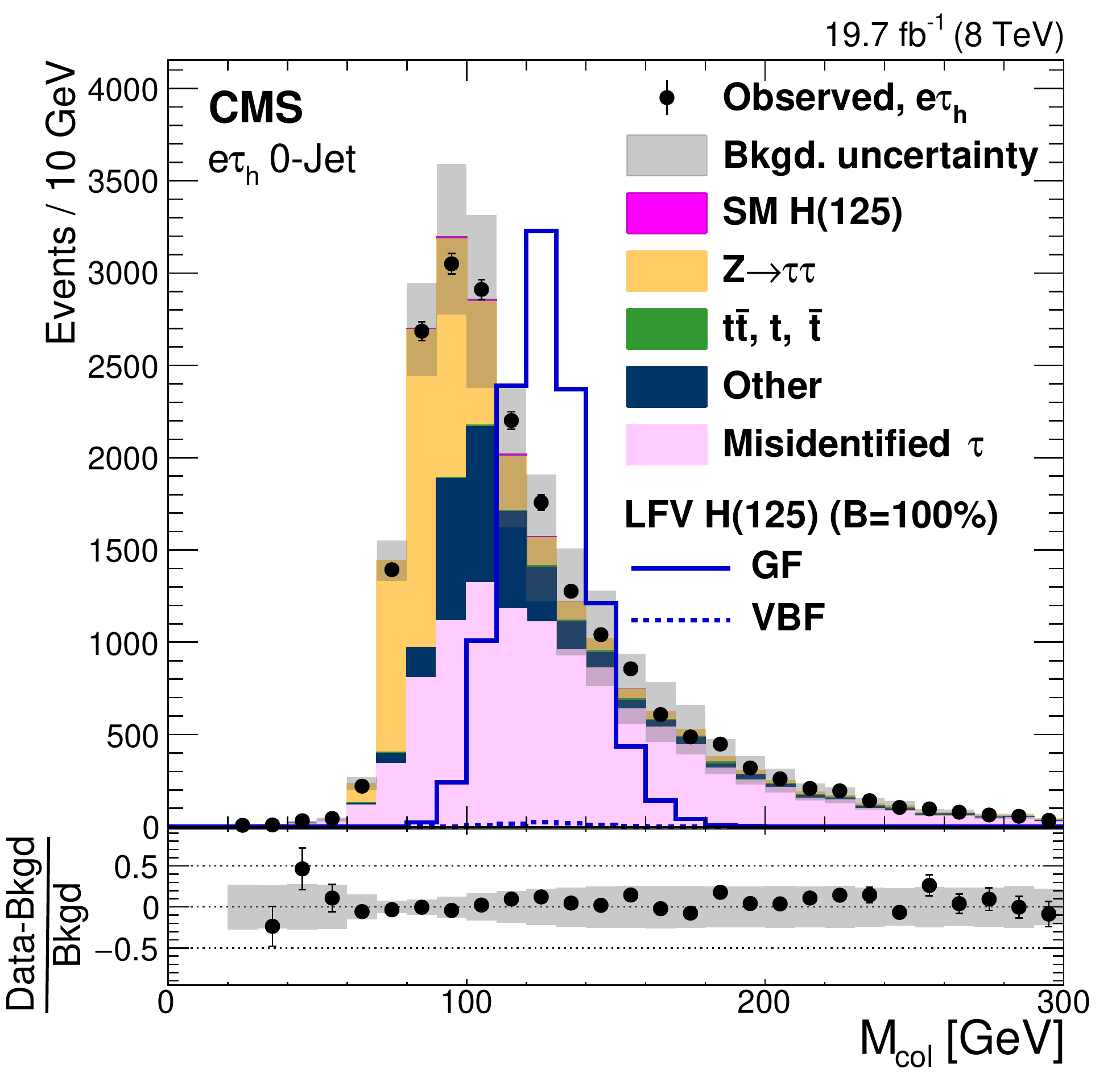}
 \includegraphics[width=0.4565\textwidth]{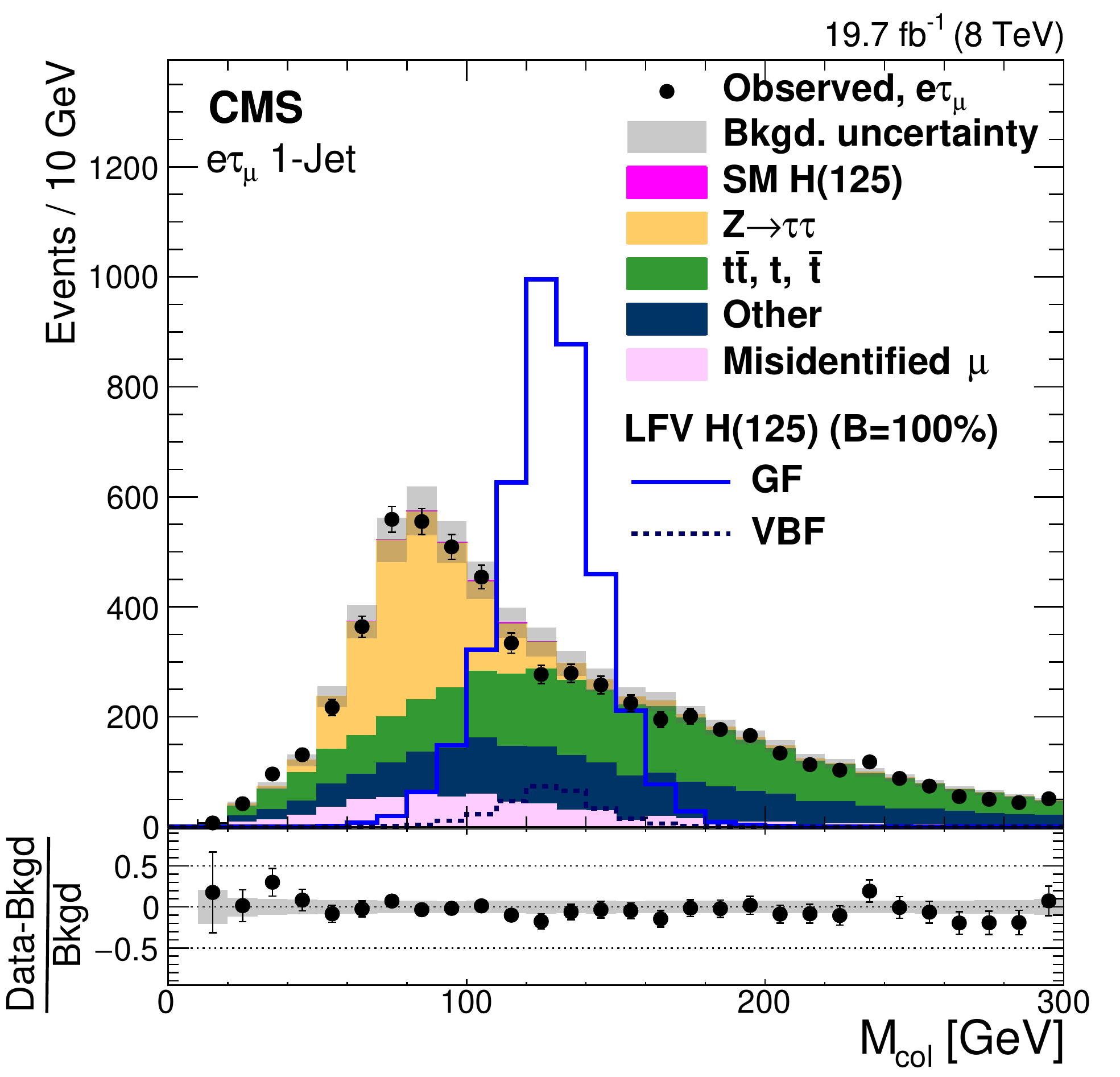}
 \includegraphics[width=0.4565\textwidth]{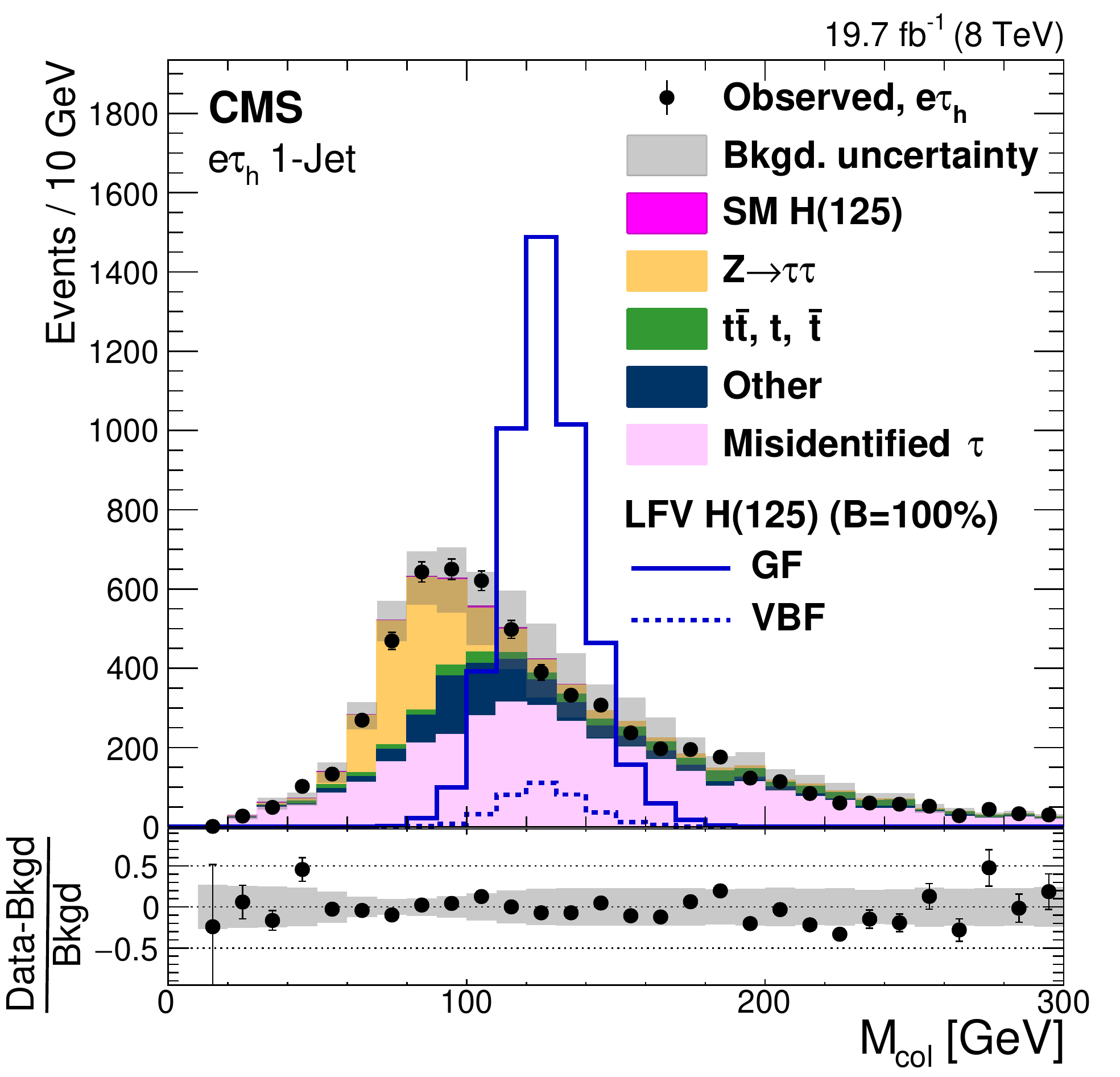}
 \includegraphics[width=0.4565\textwidth]{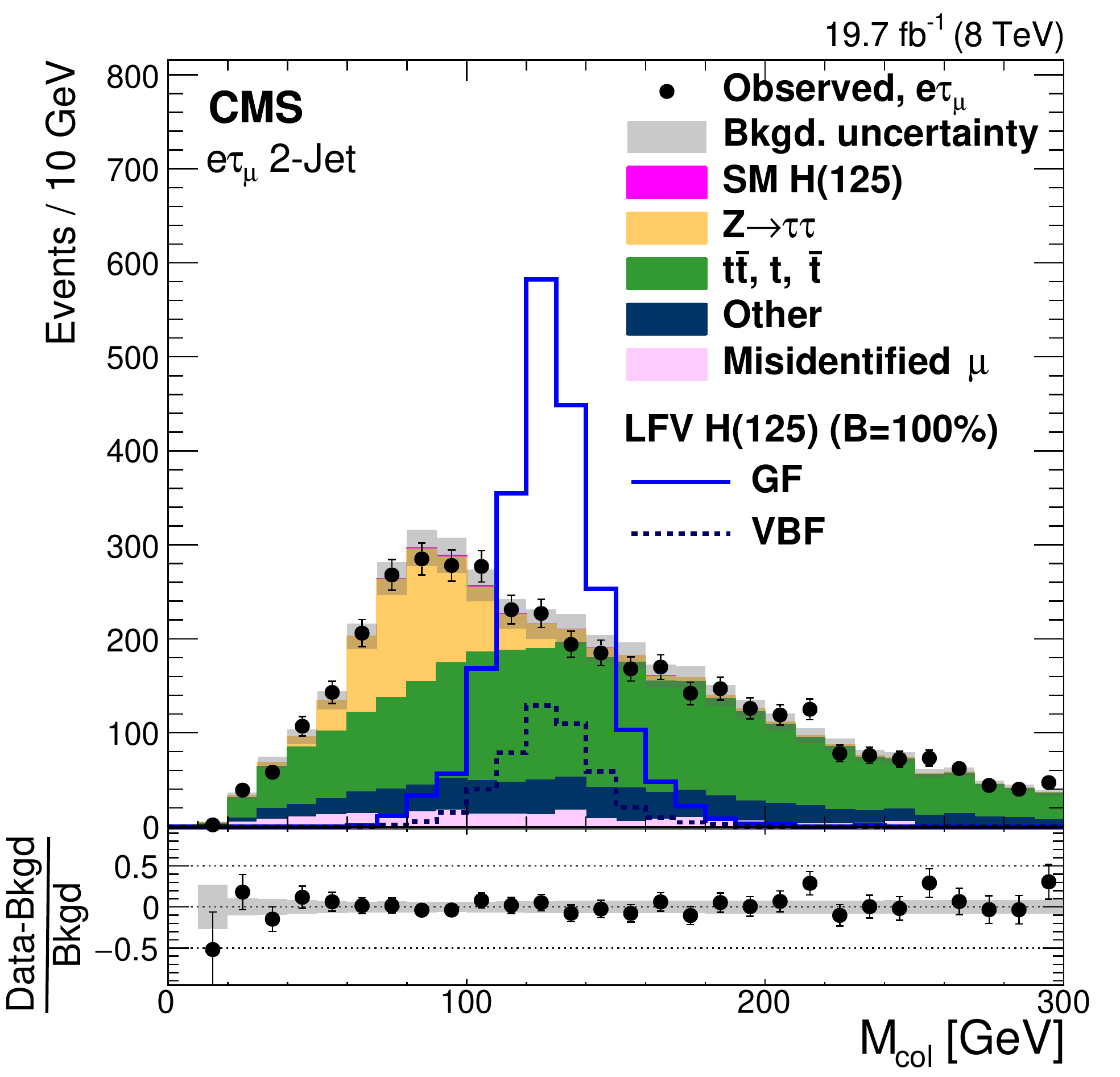}
 \includegraphics[width=0.4565\textwidth]{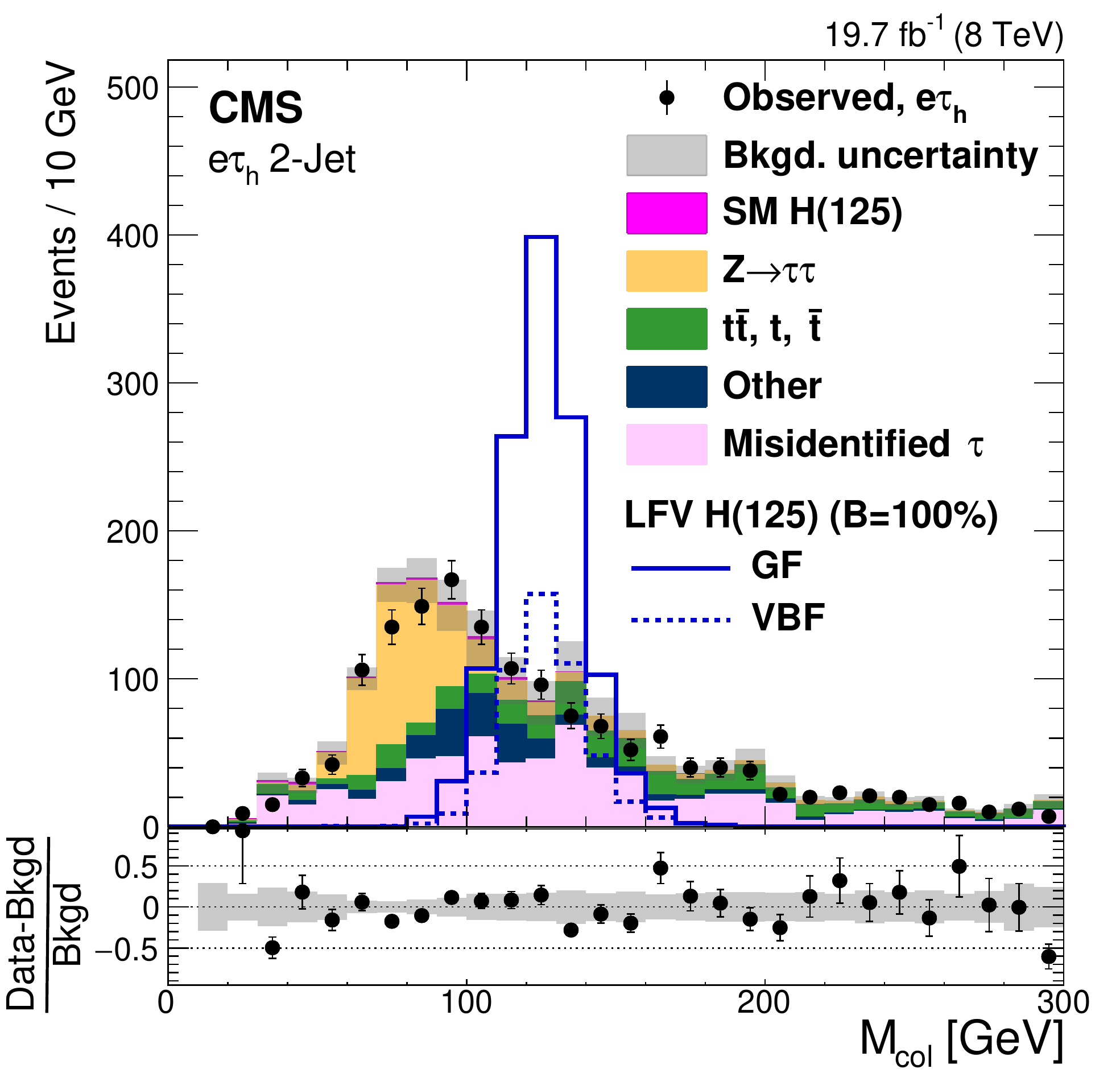}
 \caption{
Comparison of the observed collinear mass distributions with the background expectations after the loose selection requirements.
The shaded grey bands indicate the total background uncertainty.
The open histograms correspond to the expected signal distributions for $\mathcal{B}(\PH \to \Pe \Pgt )=100\%$.
The left column is  $\PH \to \Pe \Pgt_{\Pgm}$ and the right column is  $\PH \to \Pe \tauh$; the upper, middle and lower rows are the  0-jet, 1-jet and 2-jet categories, respectively.}

\label{fig:Mcol_after_presel_WITHDATA}\end{figure*}

Figure~\ref{fig:Mcol_after_presel_WITHDATA} shows the observed  $M_{\text{col}}$ distribution
and estimated backgrounds for each category and channel, after the loose selection. The
simulated signal for \mbox{$\mathcal{B}(\PH \to \Pe \Pgt )=100 \%$} is shown. The principal backgrounds are estimated with
collision data using techniques described in Section~\ref{sec:backgrounds}.  There is good agreement between
the observed distributions and the corresponding background estimations. The agreement is similar  in all of
the kinematic variables that are subsequently used to suppress backgrounds. The analysis is subsequently
performed blinded by using a fixed selection and checking the agreement between relevant observed and simulated distributions outside the sensitive region $100\GeV  < M_{\text{col}} < 150\GeV$.

\begin{table*}[hbtp]
 \centering
 \topcaption{Event selection criteria for the kinematic variables after applying loose selection requirements.}
  \label{tab:kinematicselection}
  \begin{tabular}{lccc|ccc} \hline
  Variable                     & \multicolumn{3}{c|}{$\PH \to \Pe \Pgt_{\Pgm}$}                   &     \multicolumn{3}{c}{$\PH \to \Pe \tauh$}  \\ \cline{2-7}
      [\GeV]                   &  0-jet        & 1-jet          & 2-jet         &  0-jet          & 1-jet       & 2-jet  \\ \hline
$\pt^{\Pe} $                   &  $>$50       & $>$40        &  $>$40       & $>$45         & $>$35          & $>$35    \\
$\pt^{\Pgm}$                   &   $>$15     & $>$15        &  $>$15       &  ---             & ---          & ---      \\
$\pt^{\tauh}$               &    ---         &   ---           &   ---          & $>$30          & $>$40          & $>$30    \\
$\mt^{\Pgm}$                  &    ---         &  $<$30        &  $<$40       &   ---            &  ---         & ---      \\
$\mt^{\tauh}$             &    ---         &   ---           &   ---          & $<$70          & ---          &  $<$50   \\   \hline
    [radians]                  &              \multicolumn{3}{c|}{}             &                \multicolumn{3}{c}{}       \\  \hline
$\Delta \phi_{\vec{p}_{\mathrm{T},\Pe} - \vec{p}_{\mathrm{T},\tauh}}$ &    ---         &   ---        &   ---        &  $>$2.3           & ---          & ---      \\
$\Delta \phi_{\vec{p}_{\mathrm{T},\Pgm} - \VEtmiss}$        & $<$0.8        &  $<$0.8       &   ---      &   ---           & ---          & ---      \\
$\Delta \phi_{\vec{p}_{\mathrm{T},\Pe} - \vec{p}_{\mathrm{T},\Pgm}}$ &    ---         & $>$0.5       &   ---          &   ---           &  ---         & ---      \\  \hline
  \end{tabular}
\end{table*}

Next, a set of kinematic variables is defined, and the  event selection criteria are set to maximise the
significance $\mathrm{S}/\sqrt{\mathrm{S}+\mathrm{B}}$, where S and B are the expected signal and background
event yields in the mass window $100\GeV < M_{\text{col}} < 150\GeV$. The signal event
yield corresponds to  the SM Higgs boson production  cross section at $M_{\PH}=125$\GeV with
$\mathcal{B}(\PH \to \Pe \Pgt )=1 \%$. The selection criteria for each category and channel are given in
Table~\ref{tab:kinematicselection}. The variables used are:
the lepton transverse momenta $\pt^{\ell}$ with $\ell=\Pe,\Pgm,\tauh$; azimuthal angles between the leptons
$\Delta \phi_{\ptvec^{\ell_{1}} - \ptvec^{\ell_{2}}}$;
azimuthal angle between the lepton and the \ETmiss vector  $\Delta \phi_{\ptvec^{\ell} - \VEtmiss}$; the transverse mass
$\mt^{\ell}=\sqrt{\smash[b]{2\pt^{\ell}\ETmiss(1-\cos{\Delta \phi_{\ptvec^{\ell} - \VEtmiss}})}}$.

Events in which at least one of the jets is identified
as arising from a b quark decay are vetoed using the combined secondary vertex (CSV) b-tagging
algorithm~\cite{CMS-PAS-BTV-13-001}. To enhance the VBF contribution in the 2-jet category further requirements are applied. In the $\PH \to \Pe \tauh$ channel, events in this
category are additionally required to have two jets separated by $\abs{\Delta \eta} > 2.3$  and a
dijet invariant  mass $M_{jj}>400\GeV$.  In the $\PH \to \Pe \Pgt_{\Pgm}$ channel,
the requirements are $\abs{\Delta \eta} > 3$ and $M_{jj}>200\GeV$.

After the full selection, a binned likelihood is used to fit the distributions of $M_{\text{col}}$ for
the signal and the background contributions. The modified-frequentist \CLs method~\cite{Junk,Read2}
is used to set upper bounds on the signal strength $\Pgm$, or determine a signal significance.

\subsection{Background processes}
\label{sec:backgrounds}
The contributions from the dominant background processes are estimated using collision data  while the less significant
backgrounds are estimated using simulation.  The largest backgrounds are from $\cPZ \to \Pgt \Pgt$ decays
and from $\PW\mathrm{+jets}$ and QCD multijet production. In the latter, PF objects (predominantly jets), are misidentified as leptons.

\subsubsection{\texorpdfstring{$\cPZ \to \Pgt \Pgt$ background}{ZZ to tau tau background}}
\label{sec:hmuepftauembed}
\label{sec:hmueembed}

The  $\cPZ \to \Pgt \Pgt$  background contribution is estimated using an embedding technique~\cite{CMS:2011aa,CMS-PAPERS-HIG-13-004}.
First, a sample of $\cPZ \to \Pgm \Pgm$ events is selected from collision data using the loose muon selection. The muons are then replaced
with simulated $\tau$  decays reconstructed with the PF algorithm. Thus, the key features of the event topology such as
jet multiplicity, instrumental sources of \ETmiss, and the underlying event are taken directly from collision data. Only the $\Pgt$ lepton decays are simulated. The normalization
of the sample is obtained from  simulation. The technique is validated by comparing the collinear mass distributions obtained from the $\cPZ \to \Pgt \Pgt$
simulation and  the embedding technique applied to a simulated sample of $\cPZ \to \Pgm \Pgm$ events. A shift of 2\% in the mass peak of the embedded sample
relative to simulation is observed. This shift reflects a bias in the embedding technique, which does not take the differences between muons and taus in final-state radiation of photons into account, and is corrected for.
Identification and isolation corrections obtained from the comparison are applied to the embedded sample.

\subsubsection{Misidentified lepton background}
\label{sec:fakes}
The misidentified lepton background is
estimated from collision data by defining a sample with the same
selection as the signal sample, but inverting the
isolation requirements on one of the leptons, to enrich the contribution from  \PW+jets and QCD multijets. The
probability for PF objects to be misidentified as leptons is measured
using an independent collision data set, defined below,  and this probability is applied to the background enriched sample to
compute the misidentified lepton background in the signal sample.
The technique is shown schematically in Table~\ref{tab:fakeratediagram} in which four regions
are defined including the signal~(I) and background~(III) enriched regions and two control Regions~(II~\&~IV),
defined with the same selections as Regions~I~\&~III respectively, except with leptons of the same charge.

\begin{table}[hbt]
 \centering
 {
 \renewcommand{\arraystretch}{1.1}
 \topcaption{Definition of the samples used to estimate the misidentified lepton ($\ell$) background. They
are defined by the charge of the two leptons and by the isolation requirements on each. The definition
of not-isolated differs between the two channels.}
  \label{tab:fakeratediagram}
  \begin{tabular}{c|c} \hline
\textbf{Region I}              &  \textbf{Region II}             \\ \hline
$\ell^{\pm}_{1}$(isolated)  &  $\ell^{\pm}_{1}$(isolated)             \\
$\ell^{\mp}_{2}$(isolated)  &  $\ell^{\pm}_{2}$(isolated)             \\

\hline \hline
\textbf{Region III}           &  \textbf{Region IV}             \\ \hline
$\ell^{\pm}_{1}$(isolated)  &  $\ell^{\pm}_{1}$(isolated)             \\
$\ell^{\mp}_{2}$(not-isolated )  &  $\ell^{\pm}_{2}$(not-isolated)             \\
\hline
  \end{tabular}
}
\end{table}
The misidentified electron background is negligible in the $\PH \to \Pe \Pgt_{\Pgm}$ channel  due to the high $\pt$ electron threshold.
The misidentified muon background  is estimated with Region~I defined as  the signal selection with  an isolated electron and an isolated muon of opposite charge. Region~III is defined as  the  signal selection except the muon is required not to be isolated.
\begin{figure*}[hbtp]\centering
\includegraphics[width=0.495\textwidth]{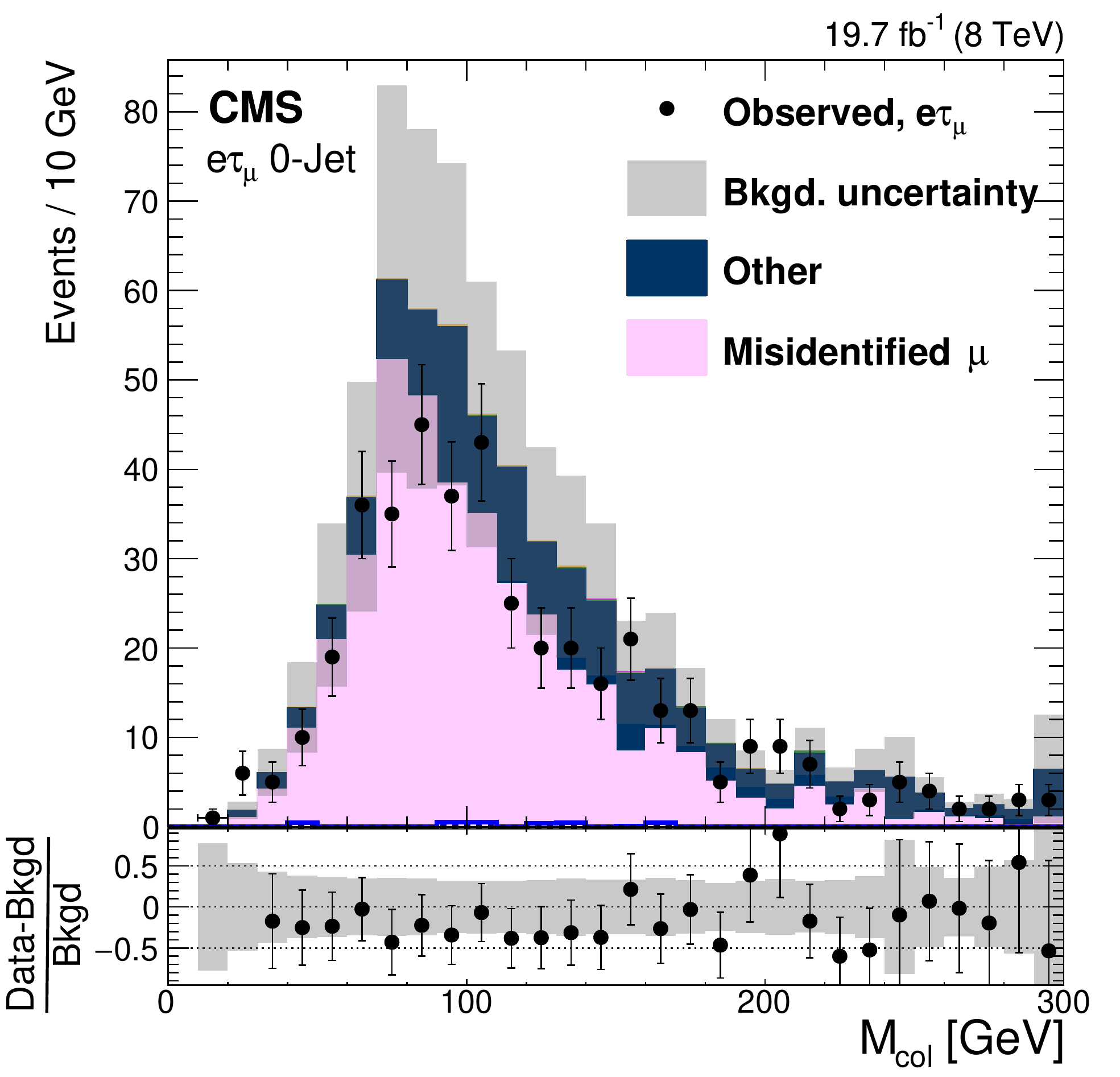}
\includegraphics[width=0.495\textwidth]{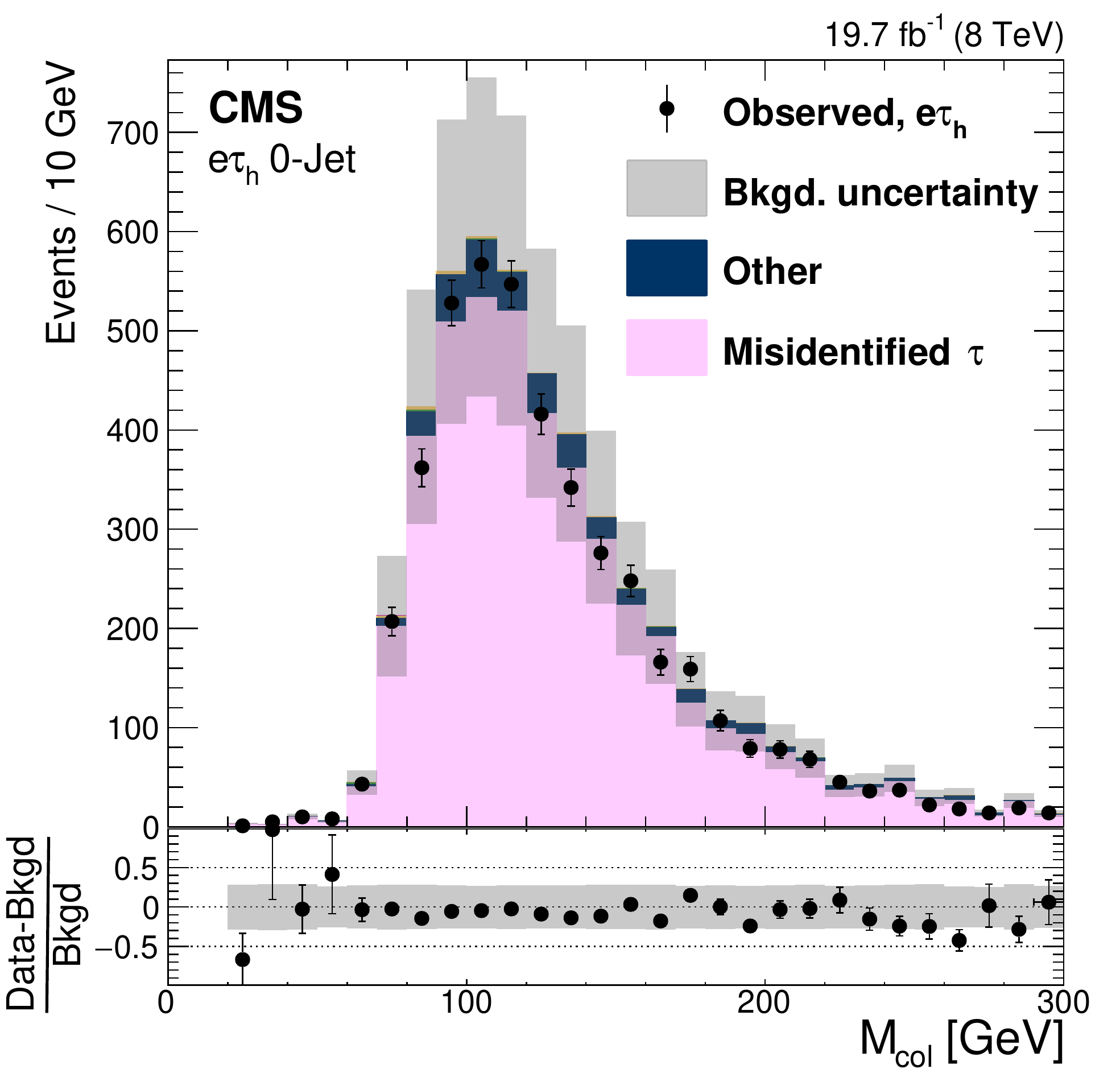}
\caption{Distributions of $M_{\text{col}}$ for Region~II. Left:  $\PH \to \Pe \Pgt_{\Pgm}$.
         Right: $\PH \to \Pe \tauh$. }
\label{fig:samesign_fakes}\end{figure*}
Small background sources  of  prompt leptons are subtracted using simulation. The
misidentified muon background in Region~I is then estimated by multiplying the event yield
in Region~III by a factor $f_{\Pgm}$, where $f_{\Pgm}$ is the ratio of isolated to nonisolated  muons.
It is computed on an independent collision data sample of $\cPZ \to \Pgm \Pgm + X$ events, where X is
an object identified as a muon, in bins of muon $\pt$ and $\eta$. In the estimation of $f_{\Pgm}$,
background sources of three prompt leptons, predominantly $\PW\cPZ$ and $\cPZ\cPZ$, are subtracted
from the  $\cPZ \to \Pgm \Pgm + X$ sample using simulation.
The technique is validated using like-sign lepton collision data in Regions~II and IV.
In Fig.~\ref{fig:samesign_fakes} (left) the event yield in Region~II is compared
to the estimate from scaling the Region~IV sample by the measured misidentification rate.
The Region~II sample is dominated by misidentified leptons but also includes  small contributions
of true leptons arising from vector boson decays,  estimated with simulated samples.

In the $\PH \to \Pe \tauh$ channel either lepton candidate can arise from a  misidentified
PF object, predominantly in  $\PW\mathrm{+jets}$ and QCD multijet events, but also from $\cPZ \to \Pe \Pe \mathrm{+jets}$
and $\ttbar$ production. The misidentification rates $f_{\Pgt}$ and $f_{\Pe}$ are defined as the fraction of
loosely isolated $\tauh$ or electron candidates that also pass a tight isolation requirement.
This is  measured in  $\cPZ \to \Pe \Pe +\mathrm{X}$ collision data events, where X is
an object identified as a $\tauh$ or  $\Pe$.
The misidentified $\tauh$ contribution is estimated  with Region~I defined as the signal selection.
Region~III is the signal selection except the $\tauh$ is required to have loose and not
tight isolation.  The misidentified $\tauh$ lepton background in Region~I is then estimated by multiplying
the event yield in Region~III by a factor $f_{\Pgt}/(1-f_{\Pgt})$.  The same procedure is used to
estimate the misidentified electron background by defining Region~I as the signal selection and Region~III as
the signal selection but with  a loose and not tight isolated electron, and scaling by $f_{\Pe}/(1-f_{\Pe})$.
To avoid double counting, the event yield in Region~III, multiplied by a factor $f_{\Pe}/(1-f_{\Pe})\times f_{\Pgt}/(1-f_{\Pgt})$, is subtracted from the sum of misidentified electrons and taus.
The procedure is validated with the like-sign $\Pe\Pgt$ samples.
Figure~\ref{fig:samesign_fakes} (right) shows the collision data  in Region~II compared to the estimate derived
from Region~IV.
The method assumes that the misidentification rate in $\cPZ \to \Pe \Pe +\mathrm{X}$ events is the same as
in the $\PW\mathrm{+jets}$ and QCD processes. To check this assumption, the misidentification rates are also measured in a collision data control sample of jets coming from QCD processes and found to be consistent. This sample
is the same $\cPZ \to \Pe \Pe +\mathrm{X}$ sample as above but with one of the electron candidates required to be not isolated and the \pt threshold lowered.

\subsubsection{Other backgrounds}
The leptonic decay of W bosons from  $\ttbar$ pairs produces opposite
sign dileptons and \ETmiss. This background is estimated using simulated $\ttbar$ events to compute the
$M_{\text{col}}$ distribution and a collision data control region for
normalization. The control region is the  2-jet selection described in Section~\ref{etauEvtSel}, including the VBF requirements, with the additional requirement that  at least one of the jets is b-tagged in order to enhance the $\ttbar$ contribution.
Other smaller backgrounds enter from SM Higgs boson production ($\PH \to \Pgt\Pgt$), $\PW\PW$, $\PW\cPZ$, $\cPZ\cPZ+\text{jets}$, $\PW\gamma^{(*)}+\text{jets}$ processes, and single top quark production. Each of these is estimated using simulation~\cite{CMS-PAPERS-HIG-13-004}.

\subsection{Systematic uncertainties}
Systematic uncertainties are implemented as nuisance parameters in the signal and background fit to determine
the scale of their effect. Some of these nuisance parameters affect only the background and signal normalizations,
while others also affect the shape of the $M_{\text{col}}$ distributions.

\subsubsection{Normalization uncertainties}
\begin{table*}[hbt]
 \centering
  \topcaption{The systematic uncertainties in the expected event yields in percentage for the $\Pe\tauh$ and $\Pe\Pgt_{\Pgm}$ channels. All uncertainties are treated as correlated between the categories, except when two values are quoted, in which case the number denoted by an asterisk is treated as uncorrelated between categories.}
  \label{tab:systematics_had}
\begin{tabular}{l|c|c|c|c|c|c} \hline
Systematic uncertainty                 & \multicolumn{3}{c}{\PH$\to \Pe\Pgt_{\Pgm}$}    & \multicolumn{3}{c}{\PH$\to \Pe\tauh$}      \\ \cline{2-7}
                                       & 0-jet       & 1-jet      & 2-jet               & 0-jet       & 1-jet        & 2-jet                 \\ \hline
Muon    trigger/ID/isolation           &  2           &  2            & 2                  & ---            & ---           & ---                        \\
Electron trigger/ID/isolation          &  3           &   3           &  3                 &  1           &   1         &  2                       \\
Efficiency of $\tauh$                & ---            & ---             & ---                  &  6.7         &  6.7        & 6.7                      \\
$\cPZ\to \Pgt \Pgt$ background & $3\oplus5^*$ & $3\oplus5^*$  & $3\oplus10^*$      & $3\oplus5^*$ & $3\oplus5^*$& $3\oplus10^*$            \\
$\cPZ\to \Pgm\Pgm, \,
\Pe\Pe$ background                     & 30           &  30           & 30                 & 30           &  30         & 30                       \\

Misidentified leptons background       & 40           &  40           & 40                 & 30           &  30         & 30                       \\
Pileup                                 & 2           &  2           & 10                 & 4           & 4          & 2                       \\
$\PW\PW,\PW\cPZ,\cPZ\cPZ\mathrm{+jets}$ background                     &  15          & 15            & 15                 & 15           &  15         & 15                       \\
$\ttbar$ background                    & 10           &  10           & $10\oplus10^*$     & 10           &  10         & $10\oplus33^*$           \\
Single top quark background                  & 25           &  25           & 25                 & 25           &  25         & 25                       \\
b-tagging veto                         & 3            &  3            & 3                  & ---            & ---           & ---                        \\
Luminosity                             & 2.6          &  2.6          & 2.6                & 2.6          &  2.6        & 2.6                      \\ \hline
\end{tabular}
\end{table*}

The values of the systematic uncertainties implemented as nuisance parameters in the signal and background fit
are summarized in Tables~\ref{tab:systematics_had} and~\ref{tab:theory_systematics}. The
uncertainties in the muon, electron and $\tauh$ selection efficiencies (trigger, identification, and isolation)
are estimated using collision data samples of $\cPZ \to \Pgm\Pgm, \Pe\Pe, \Pgt_{\Pgm}\tauh$ events~\cite{CMS:2011aa,Khachatryan:2015dfa}.
The uncertainty in the  $\cPZ \to \Pgt\Pgt$  background yield comes from the cross section uncertainty measurement (3\% \cite{Chatrchyan:2014mua}) and  from the uncertainty in
the $\Pgt$ identification efficiency when applying to the embedded technique (5-10\% uncorrelated between categories).
The uncertainties in the estimation of the misidentified lepton  rate come from the
difference in rates measured in different collision data samples (QCD multijet and $\PW\text{+jets}$).
The systematic uncertainty in the pileup modelling is evaluated by varying the total
inelastic cross section by $\pm$5\%~\cite{Chatrchyan:2012nj}.
The uncertainties in the production cross sections estimated from simulation are also
included~\cite{CMS-PAPERS-HIG-13-004}.

Uncertainties on diboson  and single top production correspond to
the uncertainties of the respective cross section
measurements~\cite{Chatrchyan2013190,Chatrchyan:2014tua}. A 10\% uncertainty from  the cross section  measurement \cite{Khachatryan:2014loa} is applied to the yield of the $\ttbar$ background. In the 2-jet categories an additional uncertainty (10\% for $\PH \to \Pe \Pgt_{\Pgm}$ and 33\% for $\PH \to \Pe \tauh$) is considered corresponding to the statistical uncertainty of the $\ttbar$ background yield.

\begin{table*}[hbtp]
 \centering
  \topcaption{Theoretical uncertainties in percentage for the Higgs boson production cross section for each
production process and category. All uncertainties are treated as fully correlated between categories except
those denoted by a negative superscript which are fully anticorrelated due to the migration of events.}
  \label{tab:theory_systematics}
  \begin{tabular}{l|l|l|l|l|l|l} \hline
Systematic uncertainty                  &  \multicolumn{3}{c|}{Gluon fusion} &  \multicolumn{3}{c}{Vector boson fusion}  \\ \cline{2-7}
                                &    0-jet  & 1-jet  & 2-jet   & 0-jet & 1-jet  & 2-jet  \\ \hline
Parton distribution function         &    $9.7$  &  $9.7$ &   $9.7$ & $3.6$  &   $3.6$  &  $3.6$  \\
Renormalization/factorization scale           &    $8$    &  $10$   &  $30^{-}$   & $4$     &   $1.5$  & $2$   \\
Underlying event/parton shower  &   $4$     & $5^{-}$   &  $10^{-}$   & $10$    &   $<$1    & $1^{-}$   \\ \hline
  \end{tabular}
\end{table*}

There are several theoretical uncertainties on the Higgs boson production
cross section that depend on the production mechanism and the
analysis category, as reported in Table~\ref{tab:theory_systematics}.
These uncertainties affect both the LFV Higgs boson and the SM Higgs boson background and are
fully correlated. The uncertainty in the parton distribution function is
evaluated by comparing the yields in each category, that span
the parameter range of three different PDF sets, CT10~\cite{Nadolsky:2008zw}, MSTW~\cite{Martin:2009iq},
NNPDF~\cite{Ball:2010de} following the  PDF4LHC~\cite{Botje:2011sn} recommendation.
The uncertainty due to the renormalization and factorization scales,  $\mu_{R}$ and  $\mu_{F}$, is estimated
by scaling up and down by a factor of two relative to their nominal values ($\mu_{R}=\mu_{F}=\MH/2$). The uncertainty in the
simulation of the underlying event and parton showers is estimated by using two different \PYTHIA tunes, AUET2 and Z2*.
All uncertainties are treated as fully correlated between categories except those denoted by a negative superscript which are fully anticorrelated due to
the migration of events.

\subsubsection{\texorpdfstring{$M_{\text{col}}$}{m[col]} shape uncertainties}
\label{sec:Hmtshape}
\begin{table}[hbtp]
 \centering
  \topcaption{Systematic uncertainties in the shape of the signal and background distributions, expressed in percentage. The systematic uncertainty and its implementation are described in the text.}
  \label{tab:shape_systematics}
  \begin{tabular}{l|c|c} \hline
Systematic Uncertainty                                 &  $\PH \to \Pe \Pgt_{\Pgm}$      &   $\PH \to \Pe \tauh$                   \\ \hline
$Z \to \Pgt \Pgt$ bias                                 &   $2$ &     ---                                          \\
$Z \to \Pe\Pe$ bias                                    &   ---         &   5                                         \\
Jet energy scale                                       &   3--7     &    3--7                                       \\
Jet energy resolution                                  &   1--10     &    1--10                                       \\
Unclustered energy scale                               &   10      &    10                                       \\
$\tauh$ energy scale                              &   ---         &    3                                         \\    \hline
  \end{tabular}
\end{table}

The systematic uncertainties that lead to a change in the shape of the  $M_{\text{col}}$ distribution
are summarized in Table~\ref{tab:shape_systematics}.
A 2\% shift in the $M_{\text{col}}$ distribution of the embedded $\cPZ \to \Pgt \Pgt$ sample used to estimate the background is observed relative to simulation. It occurs only in
the $\PH \to \Pe \Pgt_{\Pgm}$ channel as the effects of bremsstrahlung from the muon are neglected in the simulation.
The $M_{\text{col}}$ distribution is corrected by $2\pm2\%$ for this effect.
There is a systematic uncertainty of 5\% in $\Z\to \Pe\Pe$ background in the $\PH \to \Pe\tauh$ channel,  due to the mismeasured energy  of the  electron reconstructed as a $\tauh$. It causes a shift in the $M_{\text{col}}$ distribution, estimated by
comparing collision data with simulation in a control region of $\cPZ\to\Pe\Pe$ events in which one of the
two electrons that form the Z peak is also identified as a $\tauh$~\cite{Khachatryan:2015dfa}.
Corrections are applied for the jet energy scale and resolution~\cite{CMS-JME-10-011}. They are determined with dijet and
$\gamma/\cPZ\mathrm{+jets}$ collision data and the most significant uncertainty arises from the photon energy scale.
Other uncertainties such as jet fragmentation modelling, single pion response,
and uncertainties in the pileup corrections are also included. The jet energy scale
uncertainties (3--7\%) are applied as a function of $\pt$ and $\eta$, including all correlations,
to all jets in the event, propagated to the \ETmiss, and the resultant  $M_\text{col}$
distribution is used in the fit. There is also an additional uncertainty to account for the
unclustered energy scale uncertainty. The unclustered energy comes from jets below
10\GeV and PF candidates not within jets. It is also propagated to \ETmiss.
These effects cause a  shift of the $M_\text{col}$ distribution. The uncertainty in the jet energy
resolution is used to smear the jets as a function of \pt and $\eta$ and the recomputed
$M_\text{col}$ distribution is used in the fit.
A 3\% uncertainty in the  $\tauh$  energy scale is estimated by comparing $\cPZ \to \Pgt \Pgt $
events in collision data and simulation. Potential uncertainties in the shape of the misidentified lepton backgrounds are also considered.
In the \mbox{$\PH \to \Pe \Pgt_{\Pgm}$} channel the misidentified lepton rates are applied in bins of
$\pt$ and $\eta$. In the $\PH \to \Pe \tauh$ channel,
the $\tauh$  misidentification rate is found to be approximately independent of $\pt$ but
to depend on $\eta$. These rates are  all varied by one standard deviation and the differences in the
shapes are used as nuisance parameters in the fit.
Finally, the distributions used in the fit have  statistical uncertainties
in each mass bin which is included as an uncertainty that is uncorrelated between the bins.

\section{\texorpdfstring{$\PH\to \Pe \Pgm$ analysis}{H -> e mu analysis} \label{emu}}
\subsection{Event selection \label{emuEvtSel}}
\label{sec:selection}
To select $\PH\to \Pe \Pgm$ events, the trigger requirement is an electron and a muon with $\pt$ greater than 17 and 8\GeV respectively. To enhance the signal sensitivity the event sample is
divided into nine different categories according to the region of detection of the leptons and the number of jets, and a further two categories enriched in
vector boson fusion production. The resolution of the reconstructed mass of the electron muon system, $M_{\Pe\Pgm}$, depends on whether the leptons are detected in the barrel ($\abs{\eta_{\Pe}} < 1.48$, $\abs{\eta_{\Pgm}} < 0.80$)
or endcap ($1.57 < \abs{\eta_{\Pe}} < 2.50$, $0.8 < \abs{\eta_{\Pgm}} < 2.4$), while the composition and rate of backgrounds varies with the number of jets. The definition of the categories is shown in Table~\ref{tab:categories}. The two leptons are required to be isolated in all categories. Categories 0--8, which
are selected according  to the region of detection of the lepton and number of jets,  are mutually exclusive with jets required to have $\pt > 20\GeV$. To suppress
backgrounds with significant \ETmiss, such as $\PW\PW$+jets, \ETmiss is required to be less than 20, 25 or 30\GeV, depending on the category. Jets arising from b quark decays are vetoed using the  CSV discriminant to significantly reduce the \ttbar background. In the VBF categories, the two highest \pt jets are required to have $\abs{\eta} < 4.7$ and to be separated by $\abs{\eta_{j_1} - \eta_{j_2}} > 3.0$. In addition the jets are required to have
$\abs{\eta^{*}} = \abs{\eta_{\ell_1 \ell_2} - \frac{\eta_{j_1} + \eta_{j_2}}{2}}<2.5$, where $\ell=\Pe \text{ or }\Pgm$, $\eta_{\ell_1\ell_2}$ denotes the pseudorapidity of the dilepton system and $j_{1},j_{2}$ are the two jets.  The $\Delta\phi$ between the dijet system and the dilepton system is required to be greater than 2.6 rad. The VBF tight category  selection further requires that both jets have $\pt>30\GeV$ and the dijet invariant mass be $M_{j_1 j_2}>500\GeV$, while the  VBF loose category relaxes the second jet requirement to  $\pt>20\GeV$ with $M_{j_1 j_2}>250\GeV$ and is exclusive to the VBF tight category. The leptons
in both VBF categories can be in either the barrel or endcap. To avoid an event appearing in more than one category the VBF assignment is made first. Events with more than two jets are not considered.
The selection efficiency, summed over all categories, is 24\% (22\%) for the GF (VBF) production mechanism.

\begin{table*}[hbtp]
  \centering
  \topcaption{The $\PH \to \Pe \Pgm$ event selection criteria and background model for each event category. The categories are primarily defined according to whether the leptons are detected in the barrel ($\ell_B$) or endcap ($\ell_{EC}$), and the number of jets (N-jets). Requirements are also made on $\pt^{\ell}$, \ETmiss and a veto on jets arising from a b-quark decay. The background model function and order of that function are also given.}
  \label{tab:categories}
  \begin{tabular}{c|lccc|lc}
  \hline
  Category                   & Description &N-jets      & $\pt^{\ell}$ & $\ETmiss$ &    \multicolumn{2}{c}{Background model}   \\ \cline{6-7}
                             &             &            & [\GeVns{}]        & [\GeVns{}]     &    \multicolumn{1}{c}{Function} & Order   \\ \hline
0  & $\Pe_{\mathrm{B}}\Pgm_{\mathrm{B}}$        & 0 & $>$25 & $<$30 & polynomial        & 4              \\
1  & $\Pe_{\mathrm{B}}\Pgm_{\mathrm{B}}$        & 1 & $>$22 & $<$30 & polynomial        & 4              \\
2  & $\Pe_{\mathrm{B}}\Pgm_{\mathrm{B}}$        & 2 & $>$25 & $<$25 & power law         & 1              \\
3  & $\Pe_{\mathrm{B}}\Pgm_{\mathrm{EC}}$       & 0 & $>$20 & $<$30 & polynomial        & 4              \\
4  & $\Pe_{\mathrm{B}}\Pgm_{\mathrm{EC}}$       & 1 & $>$22 & $<$20 & exponential       & 1              \\
5  & $\Pe_{\mathrm{B}}\Pgm_{\mathrm{EC}}$       & 2 & $>$20 & $<$30 & exponential       & 1              \\
6  & $\Pe_\mathrm{EC}\Pgm_{\text{B or EC}}$        & 0 & $>$20 & $<$30 & polynomial        & 4              \\
7  & $\Pe_\mathrm{EC}\Pgm_{\text{B or EC}}$        & 1 & $>$22 & $<$20 & power law         & 1              \\
8  & $\Pe_\mathrm{EC}\Pgm_{\text{B or EC}}$        & 2 & $>$20 & $<$30 & polynomial        & 4              \\
9  & VBF Tight                              & 2 & $>$22 & $<$30 & exponential       & 1              \\
10 & VBF Loose                              & 2 & $>$22 & $<$25 & exponential       & 1              \\ \hline
\end{tabular}
\end{table*}

\subsection{Signal and background modelling}
\label{sec:sigbg-model}
The signal model is the sum of two Gaussian functions, determined from simulation
for each category.
The reconstructed mass resolutions depend on whether the leptons are in the barrel (B) or
endcap (EC) calorimeter and are:
2.0--2.1\GeV for $\Pe_{\mathrm{B}}\Pgm_{\mathrm{B}}$,
2.4--2.5\GeV for $\Pe_{\mathrm{B}}\Pgm_{\mathrm{EC}}$,
3.2--3.6\GeV for $\Pe_{EC}\Pgm_{\text{B or EC}}$ categories
and 2.4 (4.0)\GeV for the VBF tight (loose) categories.
The background,  modelled as either a polynomial function,
a sum of exponential functions, or a sum of power law functions is given in
Table~\ref{tab:categories} for each category. The procedure to determine
the background function follows the method described in~\cite{Chatrchyan:2013lba}.
It is designed to choose a model with sufficient parameters to accurately describe the background
while ensuring that the signal shape is not absorbed into the background function.
The background model for each category is chosen independently using this procedure.

In a first step, reference functions are selected for each type of function
(polynomial, sum of exponentials, sum of power laws).
The order of the function is chosen such that the
next higher order does not give a significantly better fit result when fit to the
observed $M_{\Pe\Pgm}$ distribution in the range $110\GeV < M_{\Pe\Pgm} < 160\GeV$.

In a second step, an ensemble of distributions is drawn from
each of the three reference background models combined with a signal
contribution corresponding to $\mathcal{B}(\PH\to\Pe\Pgm)=0.1$\%, and fitted
for signal and background with each of the three classes of functions of different orders.

On average, the signal yield extracted from the distributions using a signal plus background fit
will differ from the injected signal due to the imperfect modelling of the background.
The bias is defined as the median deviation of the fit signal event yield from the generated number of signal events.
The possible combinations of generated distributions with the fit signal plus background models
are then reduced by requiring the bias to be less than a threshold which results in less than 1\%
uncertainty in the fit signal event yield. The combination in which the fit model has the least parameters is
then selected and the fit function is used as the background model for the collision data. If there is
more than one model with the same minimal number of parameters then the one with the least bias is
selected.

\subsection{Systematic uncertainties}
\label{sec:syst}
The systematic  uncertainties are summarized in Table~\ref{tab:emusyst}.
The background is fit to the observed mass distribution  with a negligible systematic uncertainty of $<$1\% in the signal
yield arising from the choice of background model as described above.
\begin{table}
\topcaption{\label{tab:emusyst}Systematic uncertainties in percentage on the expected yield for $\PH \to \Pe \Pgm$.
Ranges are given where the uncertainty varies with production process and category. All uncertainties are treated as correlated between categories.}
\centering
\cmsTable{
\begin{tabular}{l|c}
\hline
\multicolumn{2}{c}{Experimental uncertainties} \\\hline
Background model                             &  $<$1       \\
Trigger efficiency          & 1.0             \\
Lepton identification       & 2.0             \\
Lepton energy scale         & 1.0             \\
Dilepton mass resolution   & 5.0             \\
Pileup                      & 0.7--2.3     \\
b quark jet veto efficiency            & 0.05--0.70  \\
Luminosity                  & 2.6             \\
Jet energy scale (inclusive categories)      & 0.6--22  \\
Jet energy scale (VBF categories)            & 0.1--78  \\
Jet energy resolution (inclusive categories) & 2.8--12  \\
Jet energy resolution (VBF categories)       & 0.0--49  \\
Acceptance (PDF variations) & 0.8--5.1  \\
\hline
\multicolumn{2}{c}{Theoretical uncertainties} \\
\hline\\[-2.2ex]
GF normalization/factorization scale & $^{+7.2}_{-7.8}$ \\[0.7ex]
GF parton distribution function  & $^{+7.5}_{-6.9}$ \\[0.7ex]
VBF normalization/factorization scale      & $\pm$0.2 \\
VBF parton distribution function  & $^{+2.6}_{-2.8}$ \\[0.4ex]
\hline
\end{tabular}
}
\end{table}
Correction factors are applied to the  lepton trigger, isolation, and identification efficiencies
for  each simulated signal sample to adjust for discrepancies with the collision data.
The uncertainty in the signal yield from the lepton isolation and identification corrections is 2.0\% and is estimated with the ``tag-and-probe'' method~\cite{CMS:2011aa} applied to a collision data sample of $\cPZ$ bosons decaying to lepton pairs~\cite{Khachatryan:2015hwa,ref:Muscle}. The uncertainties in the lepton energy scale and the
dilepton mass resolution are taken from the $\PH \to \cPZ\cPZ$ analysis~\cite{HIG-13-002}.
The systematic uncertainty in the pileup modelling is evaluated by varying the total
inelastic cross section by $\pm$5\%~\cite{Chatrchyan:2012nj}.
It varies according to the production process and category between 0.7\% and  2.3\%.
There are systematic uncertainties in the  efficiency of the b quark jet veto that also
vary with production process and category from  0.05\% to 0.7\%.
The uncertainty on the integrated luminosity is 2.6\%~\cite{CMS-PAS-LUM-13-001}.
The effects of systematic uncertainties in the jet energy scale and resolution, and the uncertainties in PDF's on
the selection efficiency are estimated as described in Section~\ref{sec:Hmtshape}
for the $\PH\to\Pe\Pgt$ channel. The largest values of these systematic uncertainties occur due to the migration of events to, or from, a category with low statistics.

The theoretical uncertainties on the Higgs boson production cross section  are also described in
Section~\ref{sec:Hmtshape}.

\section{Results \label{results}}
\subsection{\texorpdfstring{$\PH\to\Pe\Pgt$}{H -> e tau}}

\begin{figure*}[hbtp]\centering
 \includegraphics[width=0.465\textwidth]{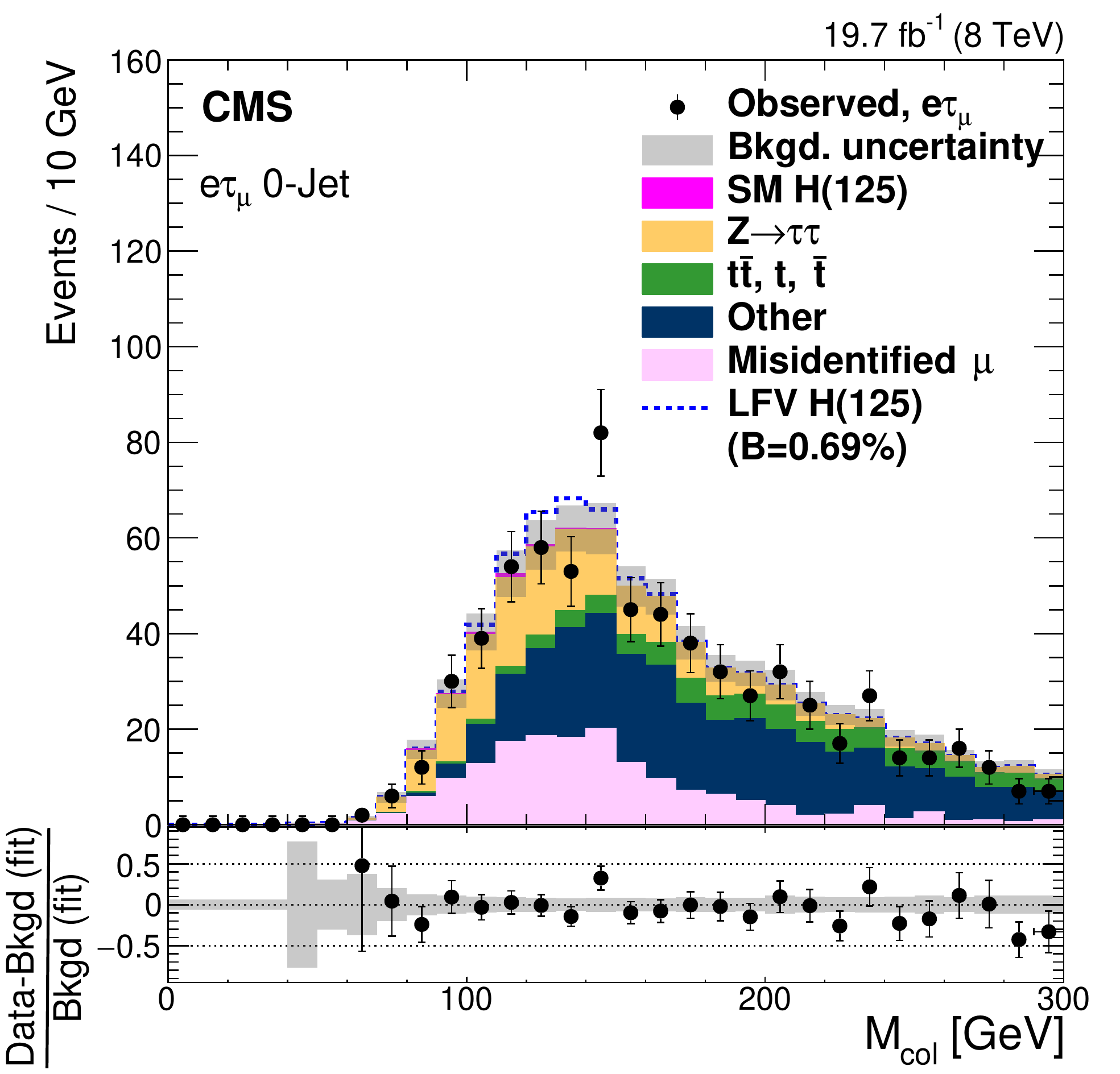}
 \includegraphics[width=0.465\textwidth]{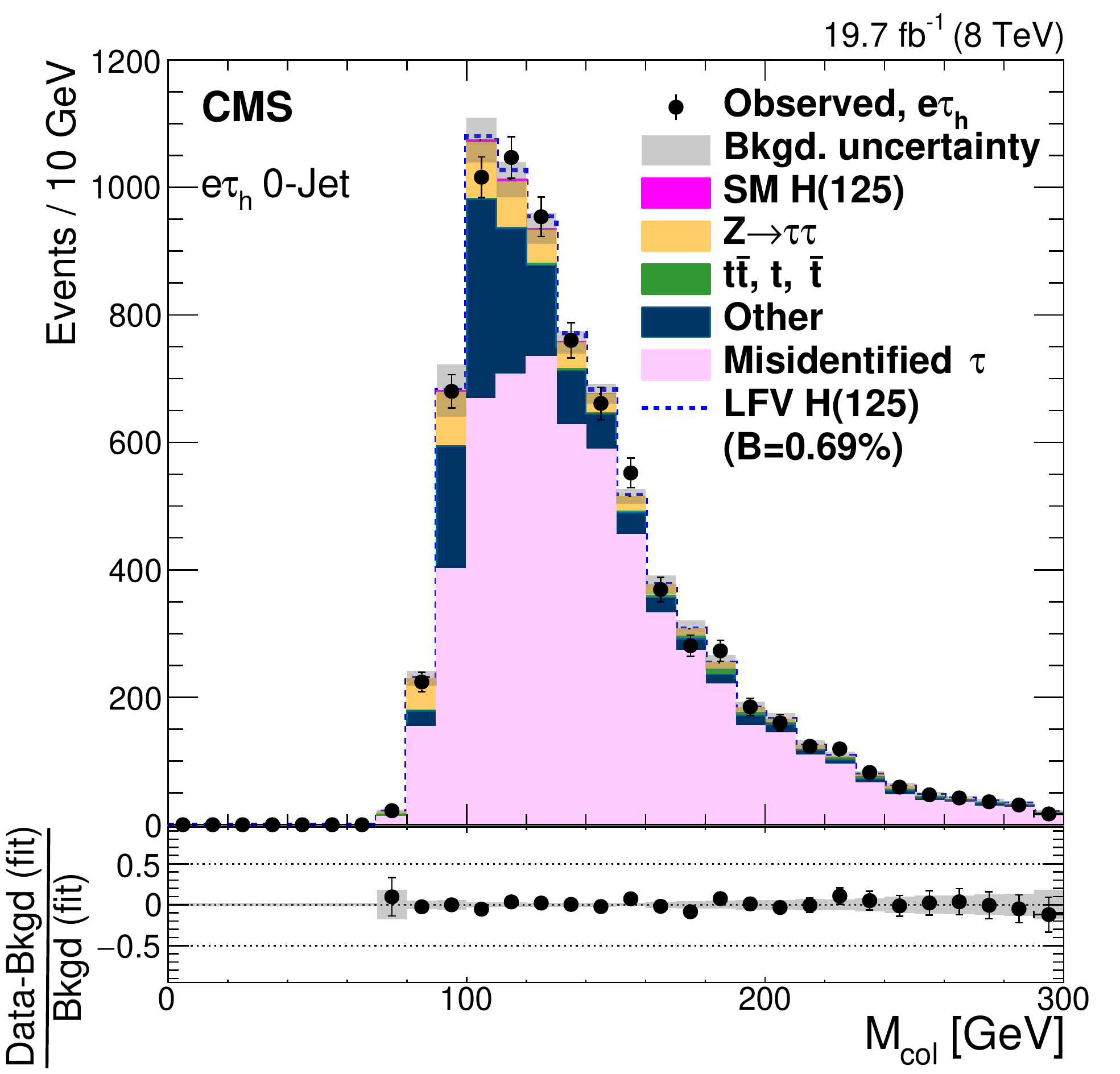}
 \includegraphics[width=0.465\textwidth]{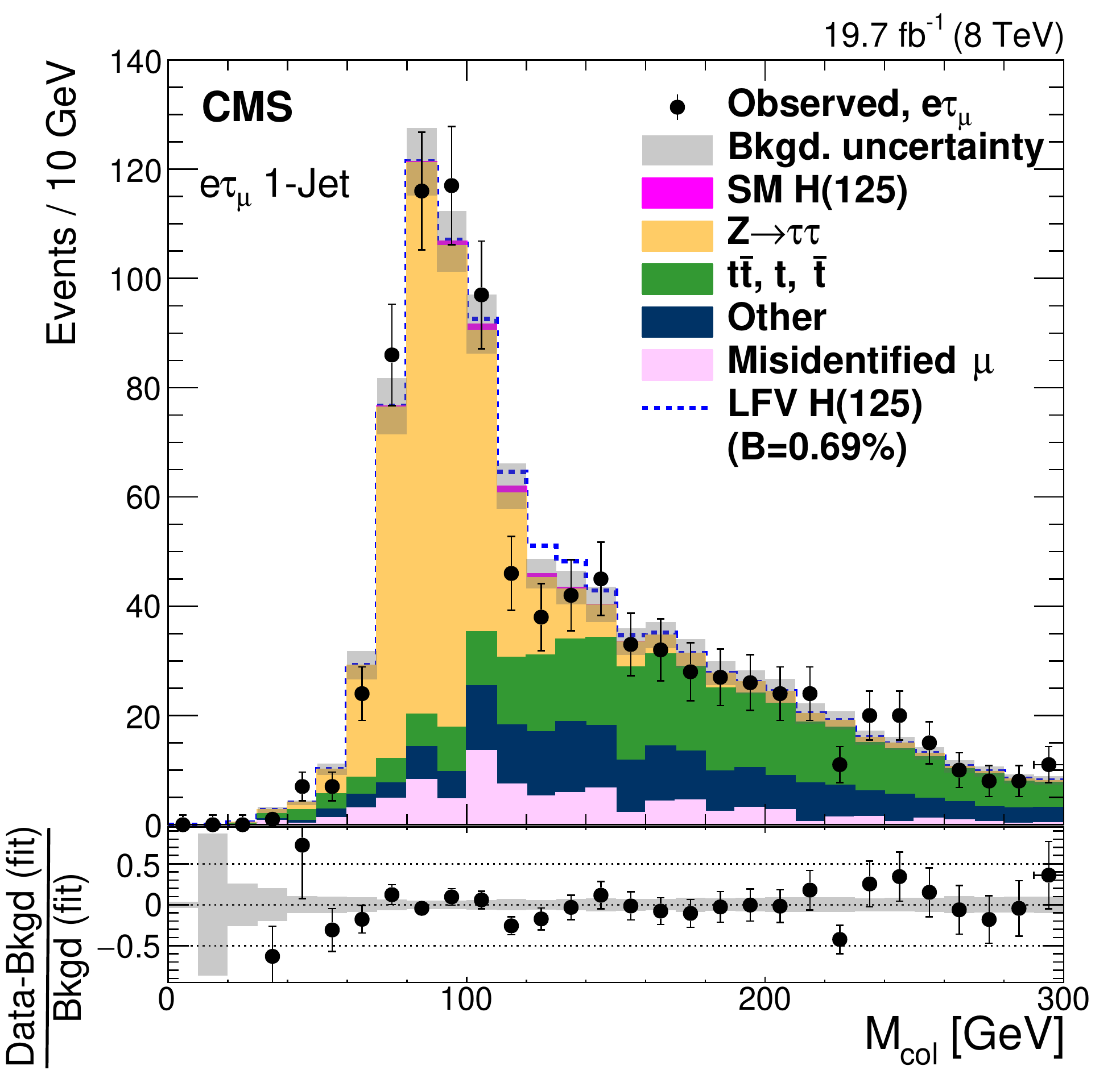}
 \includegraphics[width=0.465\textwidth]{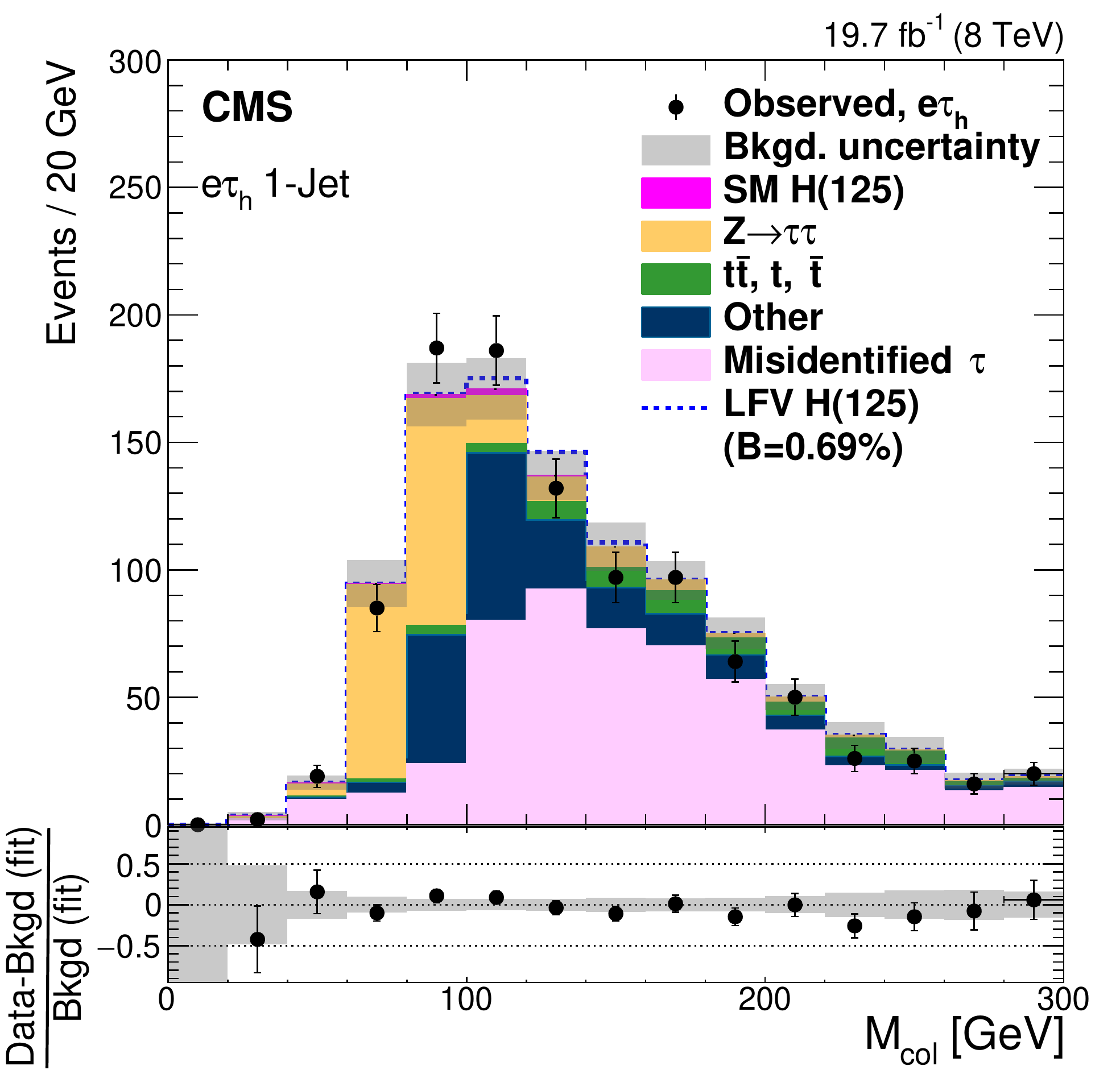}
 \includegraphics[width=0.465\textwidth]{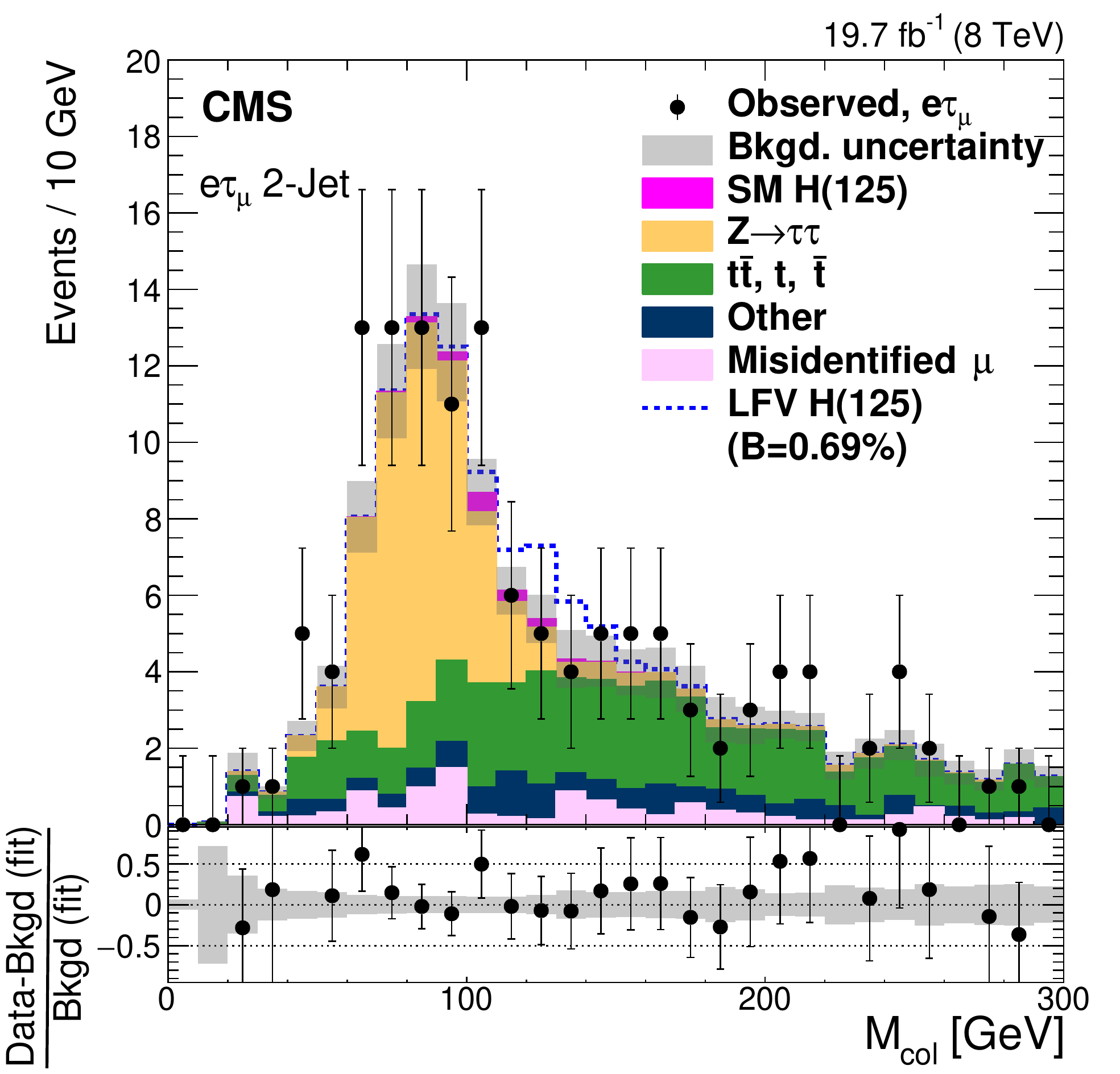}
 \includegraphics[width=0.465\textwidth]{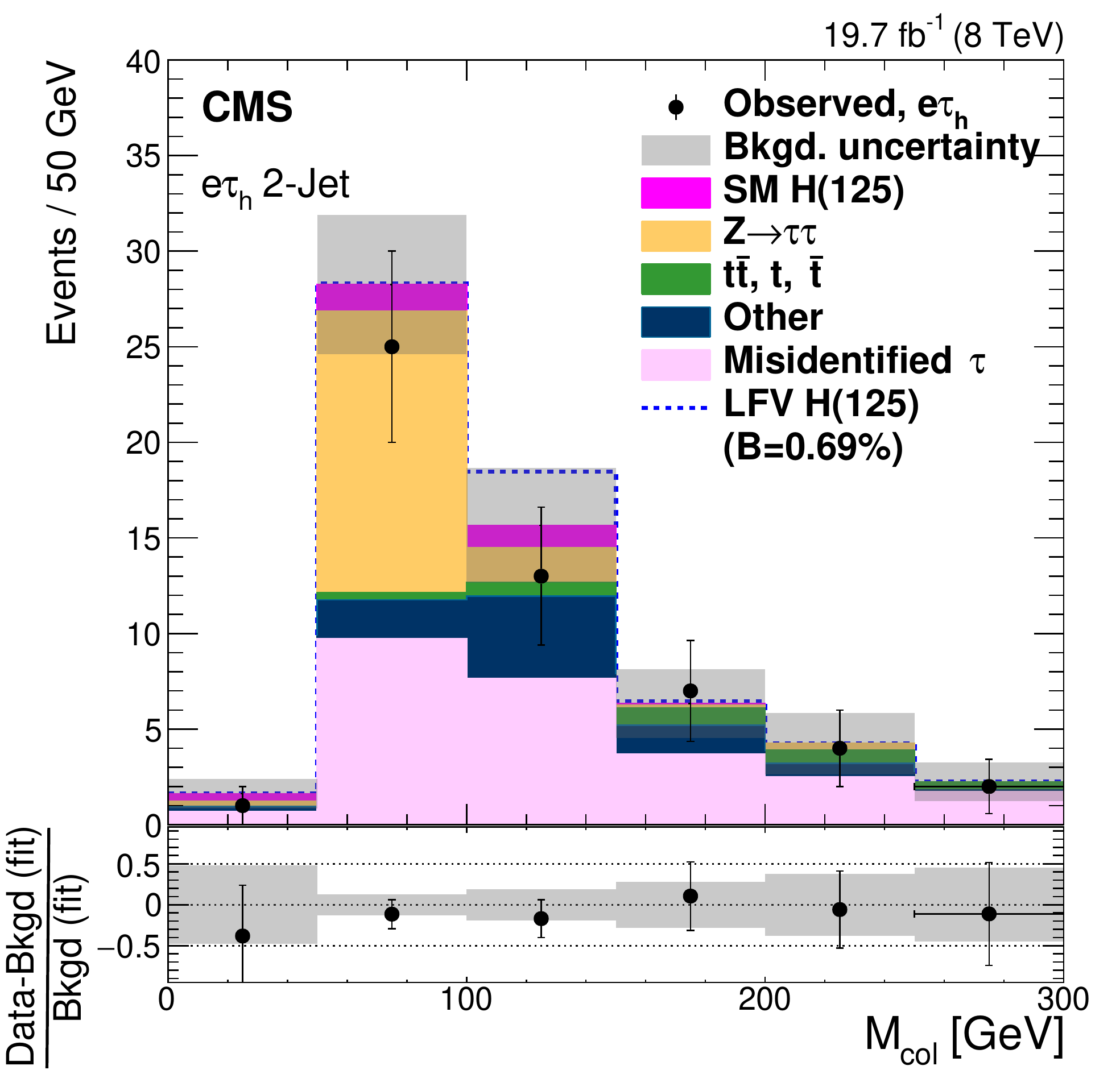}
 \caption{Comparison of the observed collinear mass distributions with the background expectations after the fit.  The simulated distributions for the signal are shown for the branching fraction $\mathcal{B}(\PH \to \Pe \Pgt )=0.69\%$. The left column is  $\PH \to \Pe \Pgt_{\Pgm}$ and the right column is  $\PH \to \Pe \tauh$; the upper, middle and lower rows are the 0-jet, 1-jet and 2-jet categories, respectively.}
 \label{fig:Mcol_Postfit}\end{figure*}

\begin{figure*}[h]\centering
\includegraphics[width=0.48\textwidth]{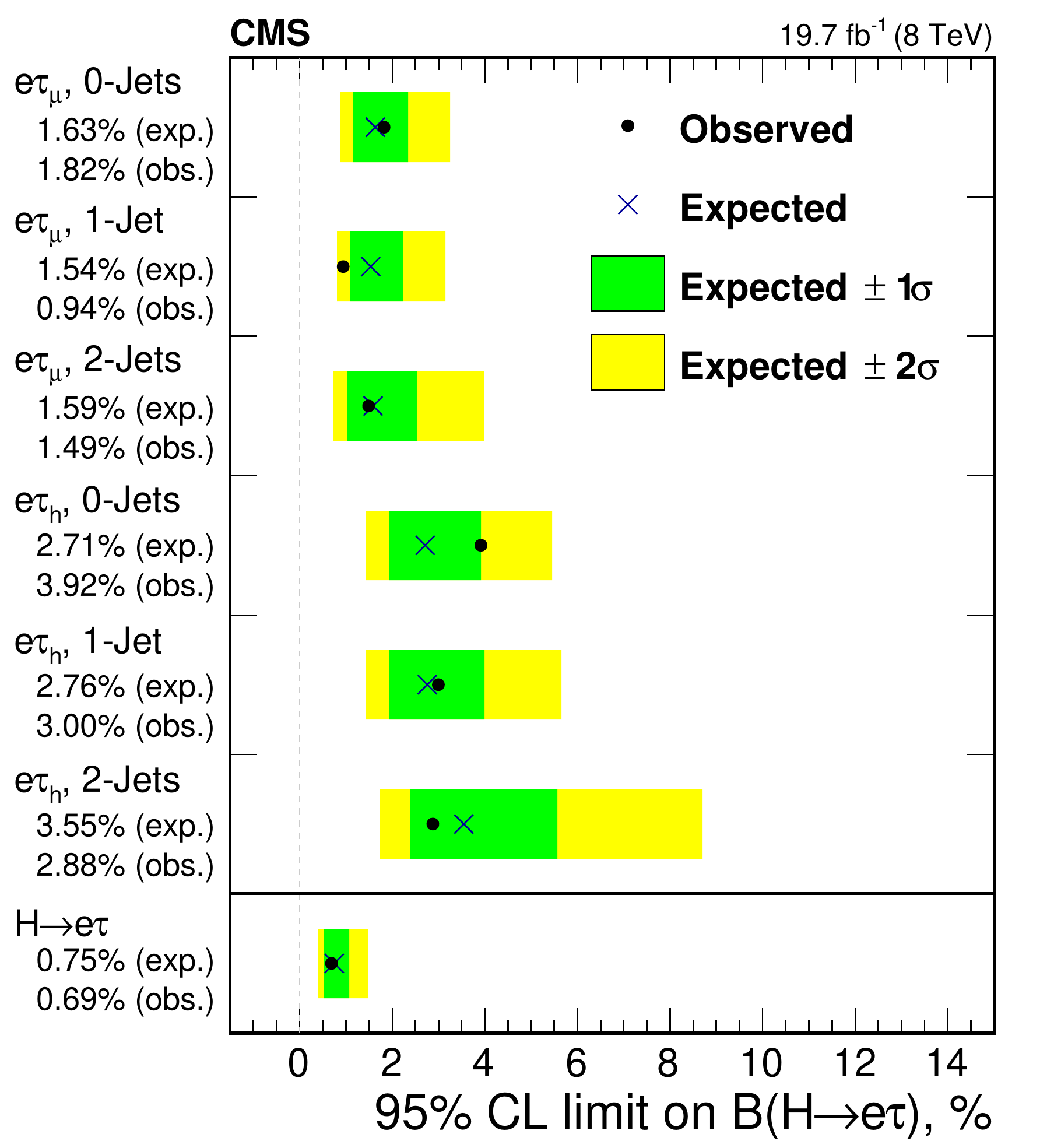}
\includegraphics[width=0.48\textwidth]{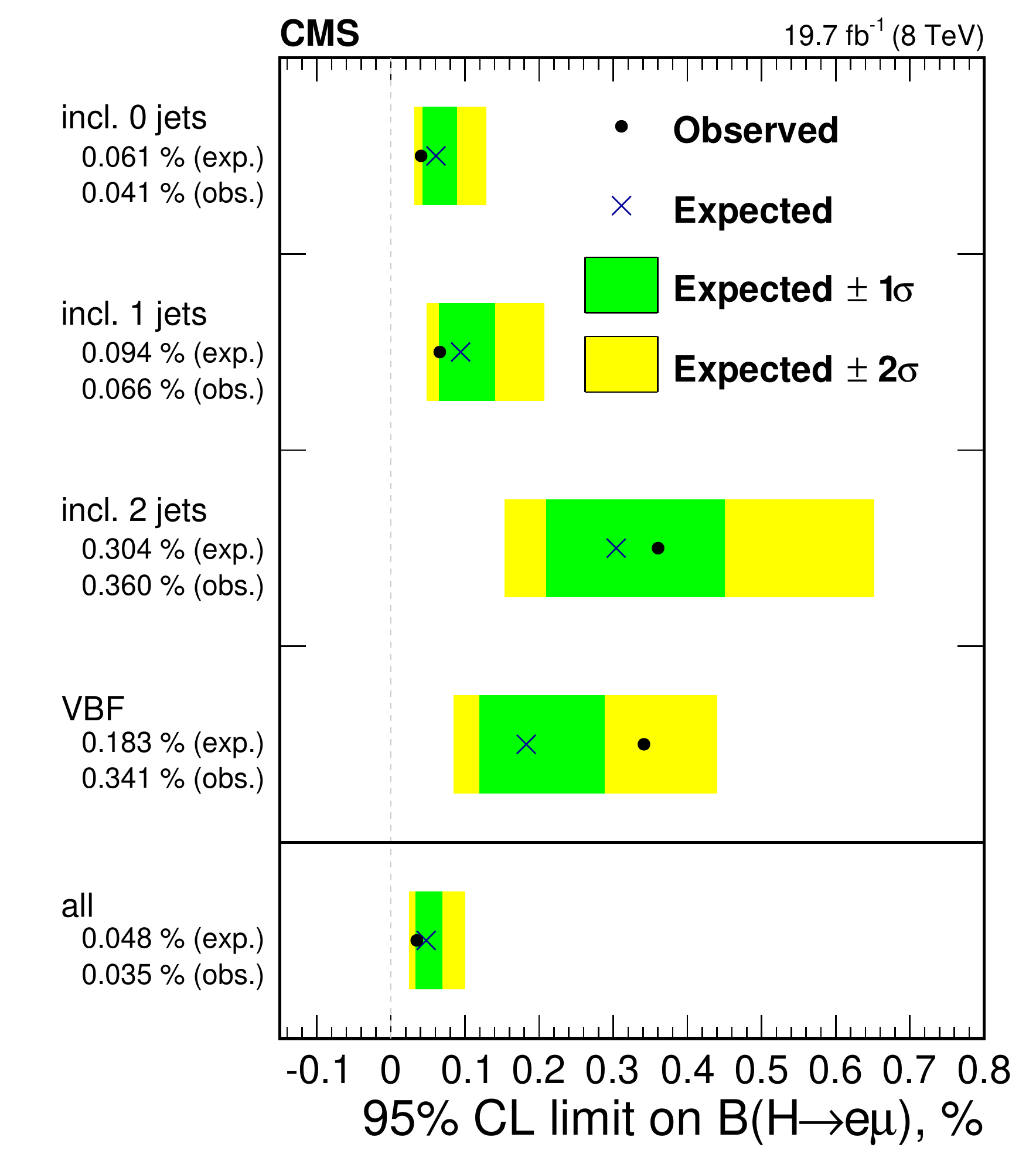}
\caption{95\% CL  upper limits by category for the LFV decays for $\MH = 125\GeV$. Left:  $\PH \to \Pe \Pgt$. Right: $\hemu$
  for categories combined by number of jets, the VBF categories combined,  and all categories combined.}
 \label{fig:limits_summary}\end{figure*}

The distributions of the fitted signal and background contributions, after the full selection,
are shown in  Fig.~\ref{fig:Mcol_Postfit} and the  corresponding event yields in the mass range
$100\GeV < M_{\text{col}} < 150\GeV$ are given in
Table~\ref{tab:EventYieldTable_100_to_150}. There is no evidence of a signal. Table~\ref{tab:expected_limits} shows the expected and
observed 95\% CL mean upper limits on $\mathcal{B}(\PH\to\Pe\Pgt)$ which are summarized in
Fig.~\ref{fig:limits_summary} for the individual categories in the  $\Pe\Pgt_{\Pgm}$ and $\Pe\tauh$ channels and for the combination.
The combined observed (expected) upper limit on $\mathcal{B}(\PH\to\Pe\Pgt)$ is 0.69\,(0.75)\% at 95\% CL.
~\cite{Junk,Read2,ATLAS:2011tau}.
\begin{table*}[h]
\centering
\topcaption{\label{tab:EventYieldTable_100_to_150} Event yields in the signal region,  $100\GeV < M_{\text{col}} < 150\GeV$, after fitting for signal and background for the $\PH\to \Pe\Pgt$ channel, normalized to an integrated luminosity of 19.7\fbinv. The LFV Higgs boson signal is the expected yield for $\mathcal{B}(\PH\to \Pe\Pgt)=0.69\%$ assuming the SM Higgs boson production cross section. }
\begin{tabular}{lccc|ccc}\hline
              & \multicolumn{3}{c|}{$\PH \to \Pe \Pgt_{\Pgm}$}                   &     \multicolumn{3}{c}{$\PH \to \Pe \tauh$}               \\ \cline{2-7}
Jet category:                          &         0-jet      & 1-jet            & 2-jet            &         0-jet    & 1-jet          & 2-jet          \\\hline
Misidentified leptons                  & 85.2$\pm$5.9       &  38.1$\pm$3.9    &  2.1$\pm$0.7     & 3366$\pm$25      & 223$\pm$11     &  8.7$\pm$ 2.2  \\
$Z\to \Pe\Pe, \Pgm\Pgm$        &  2.3$\pm$0.6       &   5.4$\pm$0.5    &  ---               & 714$\pm$30       & 85$\pm$4       &  3.2$\pm$ 0.2  \\
$Z\to \tau\tau$                & 84.7$\pm$2.1       & 113.3$\pm$4.2    &  8.5$\pm$0.6     & 270$\pm$10       & 32$\pm$3       &  1.6$\pm$ 0.3  \\
$\ttbar, \PQt, \PAQt$ & 13.8$\pm$0.3       &  69.4$\pm$2.3    &  12.7$\pm$0.8    &  10$\pm$2        &  13$\pm$2      &  0.5$\pm$ 0.2   \\
$ \cPZ\cPZ, \PW\cPZ, \PW\PW$                     & 83.0$\pm$2.7       &  51.7$\pm$2.0    &  3.6$\pm$0.4     & 53$\pm$2         &  6$\pm$1       &  0.3$\pm$ 0.1   \\
$W\gamma(^{*})$                        &  2.2$\pm$1.0       &   1.2$\pm$0.6    & ---                &  ---               &      ---         &      ---          \\
SM H  background                       &  2.3$\pm$0.3       &  3.6$\pm$0.4     &  1.1$\pm$0.2     & 12$\pm$1         &  3$\pm$1       &  1.0$\pm$ 0.1   \\
Sum of background                      & 273.5$\pm$6.1      & 282.0$\pm$6.0    & 28.1$\pm$1.3     & 4425$\pm$28      & 363$\pm$11     &  15.3$\pm$2.3   \\\hline
Observed                               & 286                & 268              & 33               & 4438             & 375            & 13            \\ \hline
LFV H  signal                          & 23.1$\pm$1.6       & 16.0$\pm$1.2     &  5.9$\pm$1.0     & 61$\pm$4        & 15$\pm$1       &  2.8$\pm$0.5    \\\hline
\end{tabular}
\end{table*}

\begin{table}[hbtp]
 \centering
  \topcaption{The expected and observed upper limits at 95\% CL, and best fit values for the branching fractions $\mathcal{B}(\PH \to \Pe \Pgt)$ for different
    jet categories and analysis channels.
    The asymmetric one standard-deviation uncertainties around the expected limits are shown in parentheses.}
  \label{tab:expected_limits}
  {
  \begin{tabular}{l|l|l|l} \hline
                     &  0-jet  & 1-jet  &  2-jet  \\ \hline
\multicolumn{4}{c}{Expected limits at 95\% CL  (\%)}\\  \hline \\[-2.2ex]
        $\Pe\Pgt_{\Pgm}$  &  ${<}1.63\left({}_{-0.44}^{+0.66}\right)$ &  ${<}1.54\left({}_{-0.47}^{+0.71}\right)$ &  ${<}1.59\left({}_{-0.55}^{+0.93}\right)$  \\[0.4ex]
        $\Pe\tauh$  &  ${<}2.71\left({}_{-0.75}^{+1.05}\right)$ & ${<}2.76\left({}_{-0.77}^{+1.07}\right)$ &  ${<}3.55\left({}_{-0.99}^{+1.38}\right)$ \\[0.4ex] \hline\\[-2.2ex]
            $\Pe\Pgt$  & \multicolumn{3}{c}{  ${<}0.75\left({}_{-0.22}^{+0.32}\right)$  } \\[0.4ex] \hline
\multicolumn{4}{c}{Observed limits at 95\% CL (\%)} \\ \hline
       $\Pe\Pgt_{\Pgm}$  &  $<$1.83  &  $<$0.94 &  $<$1.49   \\
      $\Pe\tauh$  & $<$3.92 & $<$3.00 & $<$2.88 \\ \hline
             $\Pe\Pgt$  & \multicolumn{3}{c}{  $<$0.69 }   \\ \hline
  \end{tabular}
}
\end{table}

\subsection{\texorpdfstring{$\PH\to \Pe\Pgm$}{H -> e mu}}
\label{sec:results}
\begin{figure}[htb]
\centering
\includegraphics[width=0.40\textwidth]{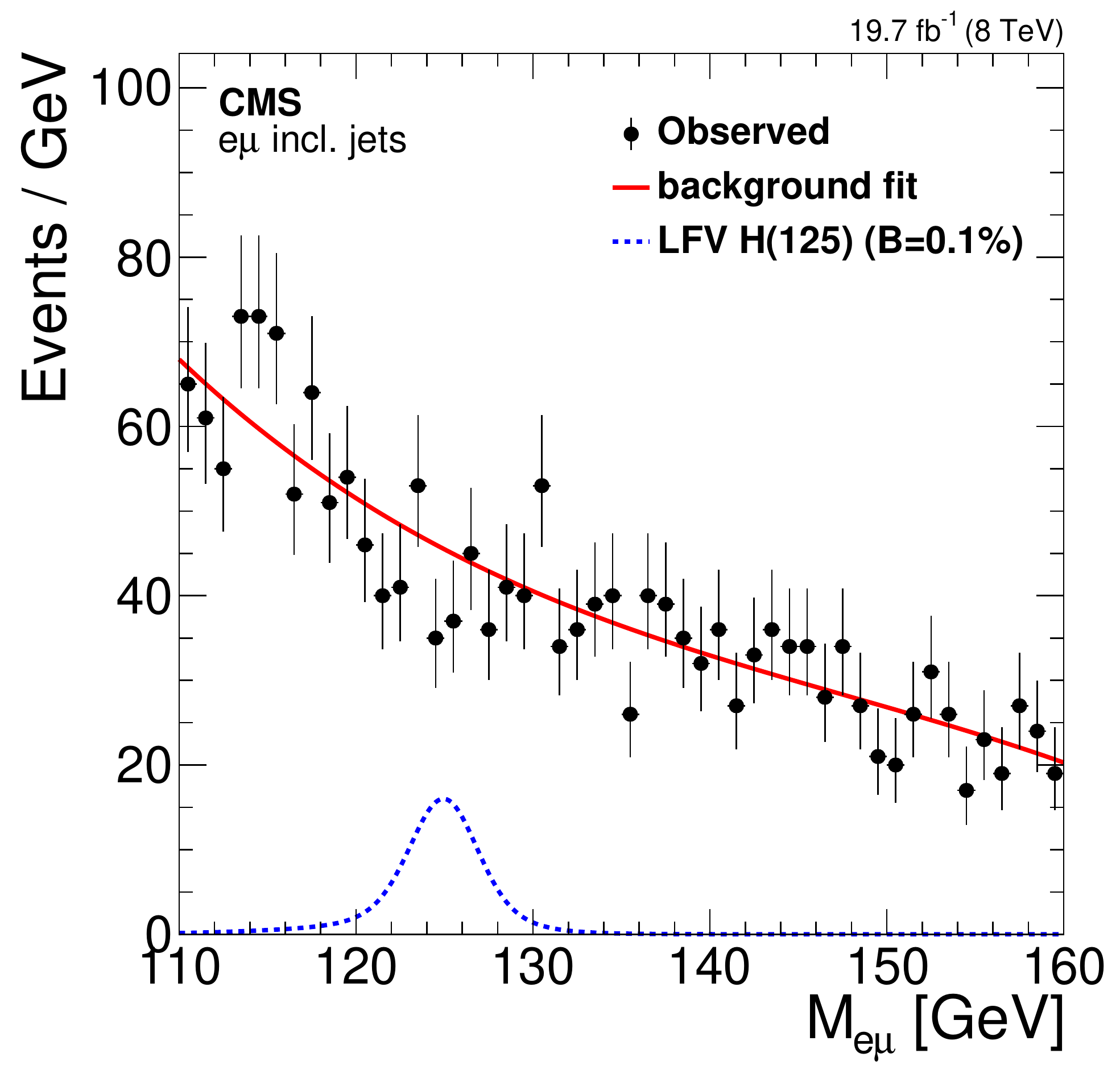}
\includegraphics[width=0.40\textwidth]{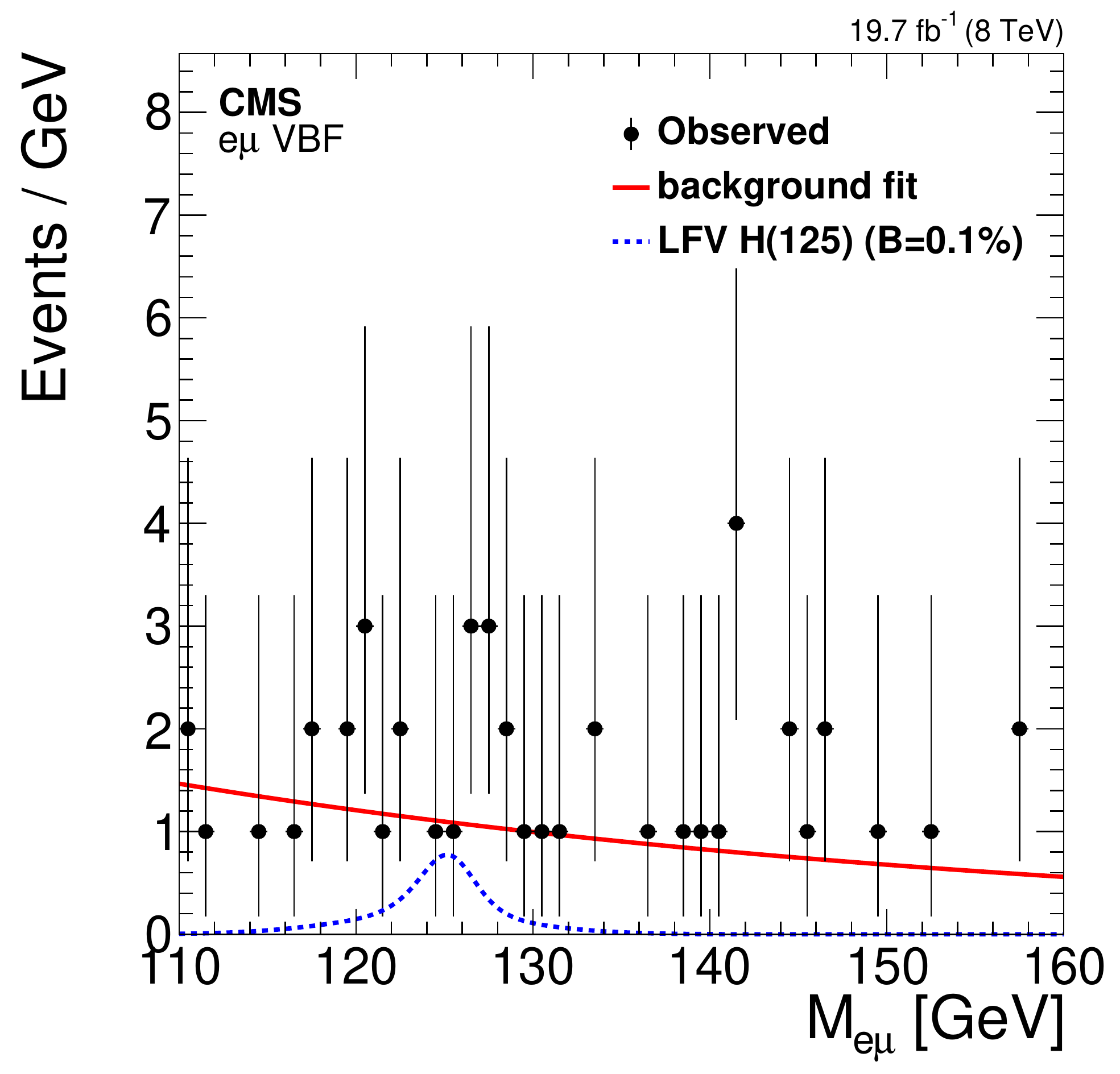}\\
\includegraphics[width=0.40\textwidth]{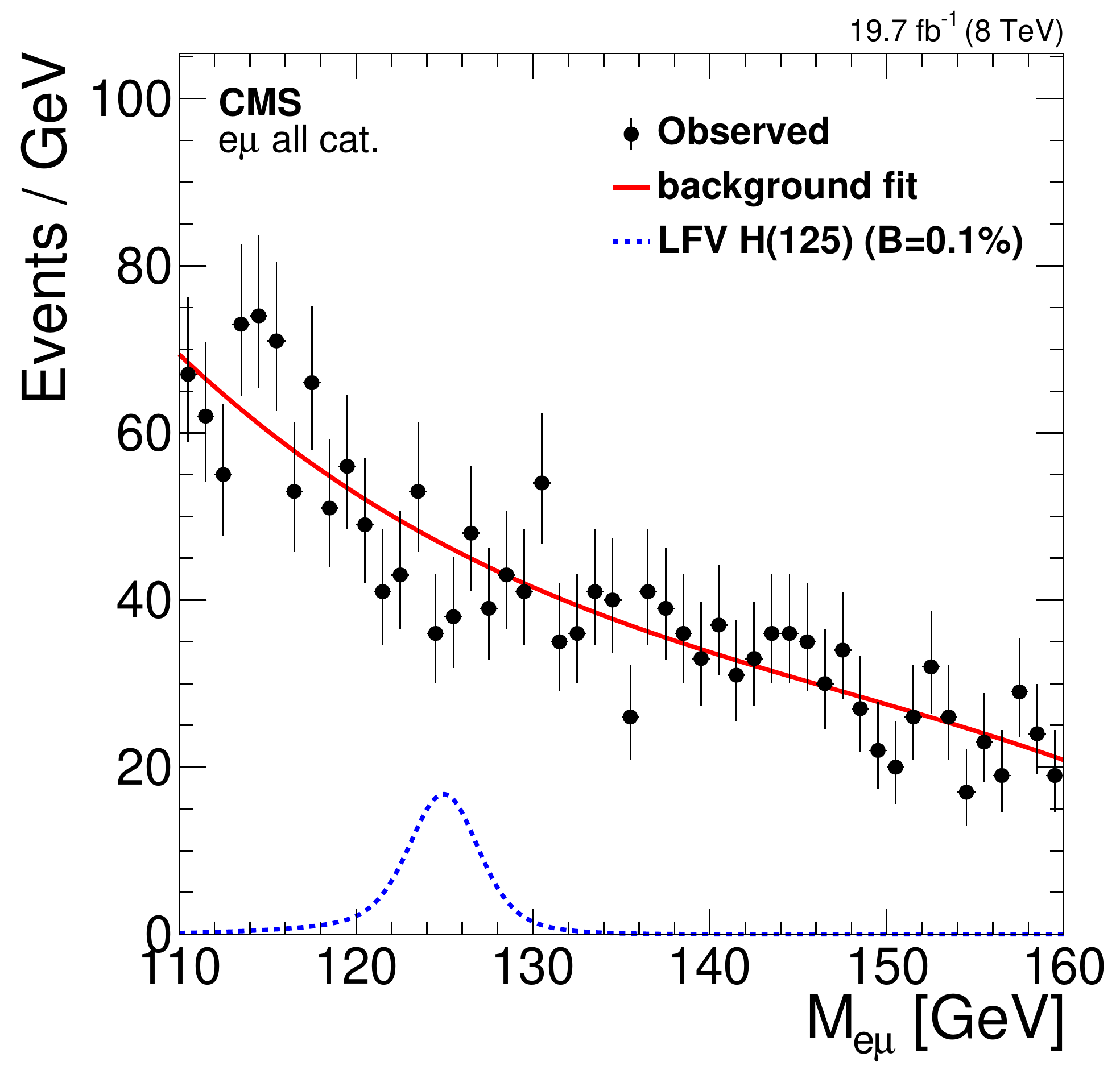}\\
\caption{Observed $\Pe\Pgm$ mass spectra (points), background fit (solid line) and signal model
(blue dashed line) for $\mathcal{B}(\hemu) = 0.1$\%. \cmsLLeft: inclusive jet categories combined (0--8). \cmsCenter: VBF jet tagged categories combined (9--10). \cmsRRight:
all categories combined.}
\label{fig:mass-inc-vbf-all}

\end{figure}
The $M_{\Pe\Pgm}$ distribution of the collision data sample, after all selection criteria, for all categories combined is shown in Fig.~\ref{fig:mass-inc-vbf-all}. Also shown are the combinations of the inclusive jet-tagged categories (0--8) and the VBF categories (9--10). The expected yields of signal ($\mathcal{B}(\hemu)=0.1$\%) and background events for $124\GeV <  M_{\Pe\Pgm} < 126$\GeV, after all the selection criteria, are given in Table~\ref{tab:hemu event yield} and compared to the collision data event yield. The contributions to the background are taken from simulation and given for information only, they are not used in the analysis. The dominant background contributions are from Drell--Yan production of $\Pgt$ lepton pairs and electroweak diboson production.
\begin{table*}[htb]
\centering
\topcaption{\label{tab:hemu event yield} Event yields in the mass window  $124\GeV <M_{\Pe\Pgm}<126$\GeV for the \hemu channel.
The expected contributions, estimated from simulation, are normalized to an integrated luminosity of 19.7\fbinv.
The LFV Higgs boson signal is the expectation for $\mathcal{B}(\hemu)=0.1\%$ assuming the SM production cross section.
Values for background processes are given for information only and are not used for the analysis.
The expected number of background events in the VBF categories obtained from simulation are associated with large uncertainties
and are therefore not quoted here; we expect $1.5\pm1.2$ events from signal and observe 2 events.
}
\begin{tabular}{lcccc}
\hline
Jet category:                   & 0-jet            & 1-jet          & 2-jet          \\\hline
Drell--Yan                       &  17.8 $\pm$ 4.2  &  6.1 $\pm$ 2.5 &  1.9 $\pm$ 1.4 \\
$\ttbar$                        &   1.4 $\pm$ 1.2  &  3.1 $\pm$ 1.8 & 14.1 $\pm$ 3.8 \\
$\PQt, \PAQt$  &   $<$1.0        &  $<$1.0       &  2.7 $\pm$ 1.6 \\
$\PW\PW$, $\PW\cPZ$, $\cPZ\cPZ$  &  21.6 $\pm$ 4.7  &  5.3 $\pm$ 2.3 &  1.9 $\pm$ 1.4 \\
SM H  background                &  $<$0.07        &  0.1 $\pm$ 0.2 &  $<$0.07      \\ \hline
Sum of backgrounds              &  40.8 $\pm$ 6.4  & 14.6 $\pm$ 3.8 & 20.7 $\pm$ 4.5 \\ \hline
Observed                        &  49              &  6             & 17             \\\hline
LFV H  signal                   &  21.2 $\pm$ 4.6  &  9.1 $\pm$ 3.0 &  2.6 $\pm$ 1.6 \\ \hline
\end{tabular}
\end{table*}
There is no signal observed. An exclusion limit on the branching fraction $\mathcal{B}(\PH\to\Pe\Pgm)$
with $M_{\text{\PH}}=125\GeV$ is derived using the \CLs asymptotic model~\cite{CLs}.
It is shown in Fig.~\ref{fig:limits_summary}
for the inclusive categories grouped by number of jets, the VBF categories,
and all categories combined.
The expected  limit is $\mathcal{B}(\PH\to\Pe\Pgm) < 0.048 \%$ at 95\% CL
and the observed limit is $\mathcal{B}(\PH\to\Pe\Pgm) < 0.035 \%$ at 95\% CL.

\subsection{\texorpdfstring{Limits on lepton flavour violating couplings }
{Limits on lepton flavour violating couplings} \label{yukawa}}
\begin{figure}[htb]
\centering
\includegraphics[width=0.47\textwidth]{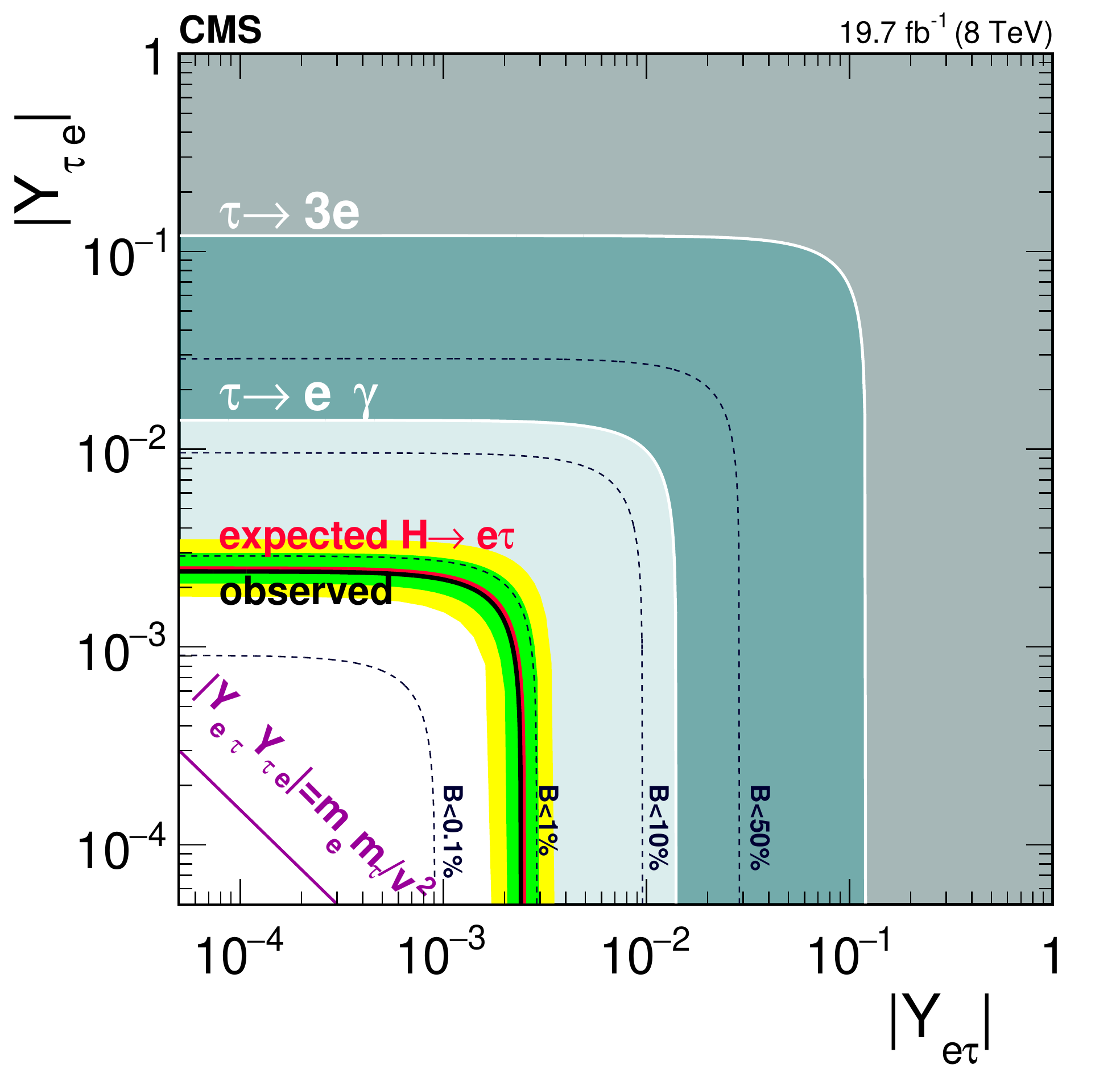}
\includegraphics[width=0.47\textwidth]{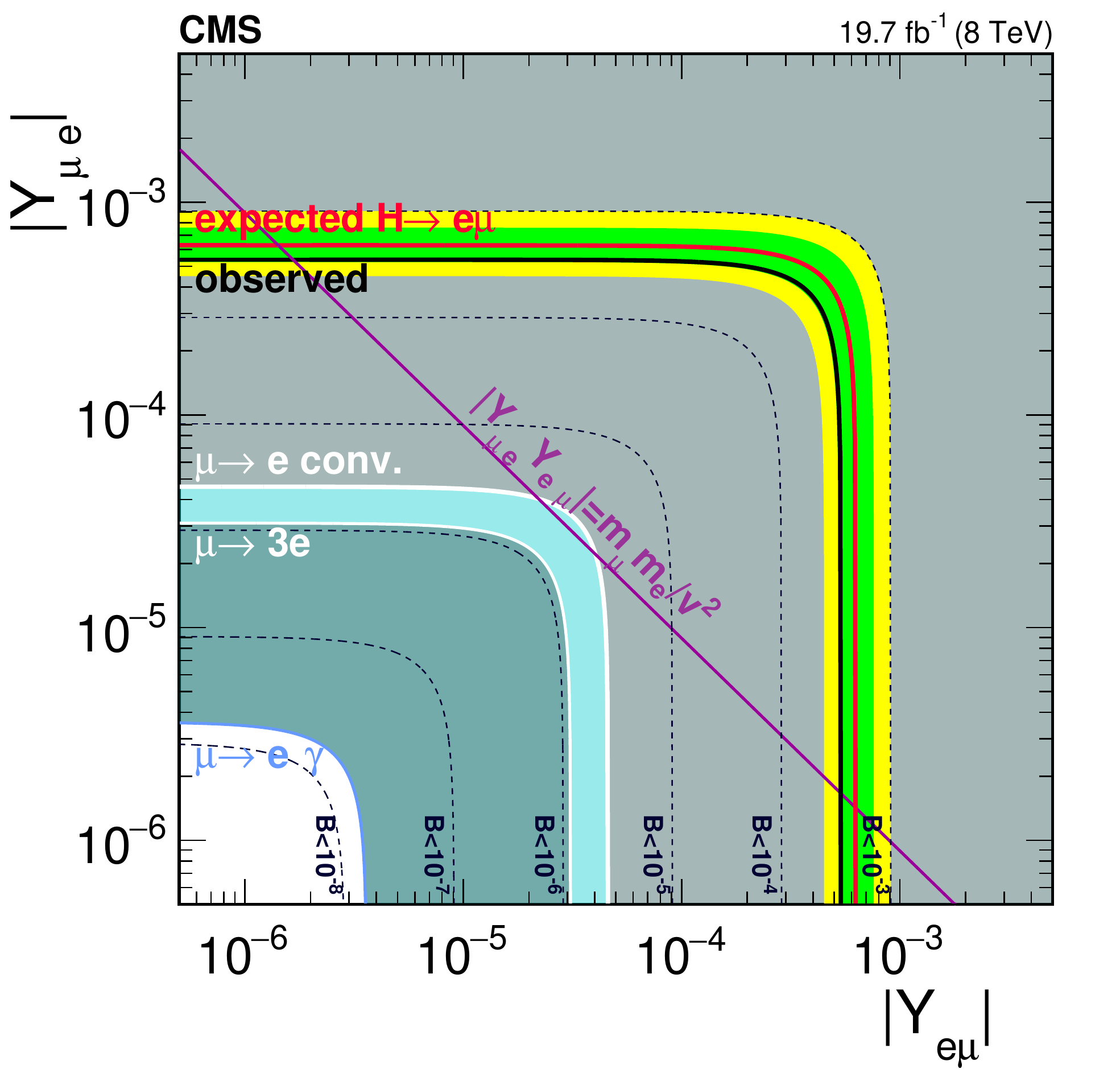}
 \caption{Constraints on the flavour violating Yukawa couplings $\abs{Y_{\Pe\Pgt}},\abs{Y_{\Pgt\Pe}}$ (\cmsLeft) and $\abs{Y_{\Pe\Pgm}},\abs{Y_{\Pgm\Pe}}$ (\cmsRight). The expected (red solid line) and observed (black solid line) limits are derived from the limits
 on $\mathcal{B}(\PH \to \Pe \Pgt )$ and $\mathcal{B}(\PH \to \Pe \Pgm )$ from the present analysis. The flavour
diagonal Yukawa couplings are approximated by their SM values. The green (yellow) band indicates the range that
is expected to contain $68\%$ ($95\%$) of all observed limit excursions.  The shaded regions in the left plot are derived constraints from null searches for $\Pgt \to 3\Pe$ (grey), $\Pgt \to \Pe \gamma$ (dark green) and the present analysis (light blue). The shaded regions in the right plot are derived constraints from null searches for $\Pgm \to \Pe \gamma$ (dark green), $\Pgm \to 3\Pe$ (light blue) and $\Pgm \to \Pe$ conversions (grey).
The purple diagonal line is the theoretical naturalness limit $Y_{ij}Y_{ji} \leq m_im_j/v^2$~\cite{Harnik:2012pb}.
}
\label{fig:yukawalimits}
\end{figure}

The constraints  on $\mathcal{B}(\PH \to \Pe \Pgt)$ and $\mathcal{B}(\PH\to\Pe\Pgm)$ can be interpreted in terms of the LFV Yukawa couplings
$\abs{Y_{\Pe\Pgt}}$, $\abs{Y_{\Pgt \Pe}}$  and $\abs{Y_{\Pe\Pgm}}$, $\abs{Y_{\Pgm\Pe}}$ respectively~\cite{Harnik:2012pb}. The LFV decays $\PH \to  \Pe\Pgt$, $\Pe\Pgm$ arise
at tree level in the Lagrangian, $L_V$, from the flavour-violating Yukawa interactions, $Y_{\ell^{\alpha}\ell^{\beta}}$, where $\ell^{\alpha},\ell^{\beta}$ denote the leptons $\Pe,\Pgm,\Pgt$,
and $\ell^{\alpha}\neq \ell^{\beta}$. The subscripts $L$ and $R$ refer to the left and right handed components of the leptons, respectively.
\ifthenelse{\boolean{cms@external}}{
\begin{multline*}
L_V \equiv - Y_{e\mu}\bar{e_L}\mu_{R}\PH - Y_{\mu e}\bar{\mu_L}e_{R}\PH 
           - Y_{e\tau}\bar{e_L}\tau_{R}\PH\\ - Y_{\tau e}\bar{\tau_L}e_{R}\PH 
           - Y_{\mu\tau}\bar{\mu_L}\tau_{R}\PH - Y_{\tau\mu}\bar{\tau_L}\mu_{R}\PH
\end{multline*}
}{
\begin{equation*}
L_V \equiv - Y_{e\mu}\bar{e_L}\mu_{R}\PH - Y_{\mu e}\bar{\mu_L}e_{R}\PH \\
           - Y_{e\tau}\bar{e_L}\tau_{R}\PH - Y_{\tau e}\bar{\tau_L}e_{R}\PH \\
           - Y_{\mu\tau}\bar{\mu_L}\tau_{R}\PH - Y_{\tau\mu}\bar{\tau_L}\mu_{R}\PH
\end{equation*}
}
The decay width $\Gamma(\PH \to \ell^{\alpha}\ell^{\beta})$  in terms of the Yukawa couplings
is given by:
\begin{equation*}
\Gamma(\PH \to \ell^{\alpha}\ell^{\beta})=\frac{M_\PH}{8\pi}(\abs{Y_{\ell^{\beta}\ell^{\alpha}}}^2 + \abs{Y_{\ell^{\alpha}\ell^{\beta}}}^2),
\end{equation*}
and the branching fraction by:
\begin{equation*}
\mathcal{B}(\PH \to \ell^{\alpha}\ell^{\beta})=\frac{\Gamma(\PH\to \ell^{\alpha}\ell^{\beta})}{\Gamma(\PH\to \ell^{\alpha}\ell^{\beta}) + \Gamma_{\mathrm{SM}}}.
\end{equation*}
The SM Higgs boson decay width is $\Gamma_{\mathrm{SM}}=4.1\MeV$ for a 125\GeV Higgs boson~\cite{Denner:2011mq}.
The 95\% CL constraints on the Yukawa couplings, derived from $\mathcal{B}(\PH \to \Pe \Pgt )<0.69\%$ and  $\mathcal{B}(\PH\to\Pe\Pgm)<0.035\%$
using the expression for the branching fraction above are:
\ifthenelse{\boolean{cms@external}}{
\begin{equation*}\begin{split}
\sqrt{\abs{Y_{\Pe\Pgt}}^{2}+\abs{Y_{\Pgt\Pe}}^{2}}<2.4\times 10^{-3}, \\
\sqrt{\abs{Y_{\Pe\Pgm}}^2 + \abs{Y_{\Pgm\Pe}}^2} < 5.4\times 10^{-4}.
\end{split}\end{equation*}
}{
\begin{equation*}
\sqrt{\abs{Y_{\Pe\Pgt}}^{2}+\abs{Y_{\Pgt\Pe}}^{2}}<2.4\times 10^{-3}, \quad
\sqrt{\abs{Y_{\Pe\Pgm}}^2 + \abs{Y_{\Pgm\Pe}}^2} < 5.4\times 10^{-4}.
\end{equation*}
}
Figures~\ref{fig:yukawalimits}  compare these results to the constraints from previous indirect
measurements. The absence of $\Pgm \to \Pe \gamma$ decays implies a  limit of $\sqrt{\abs{Y_{\Pe\Pgm}}^2 + \abs{Y_{\Pgm\Pe}}^2}< 3.6 \times 10^{-6}$~\cite{Harnik:2012pb}
assuming that flavour changing neutral currents are dominated by the Higgs boson contributions. However, this limit can be degraded by the  cancellation of
lepton flavour  violating effects from other new physics. The direct search for $\PH \to \Pe \Pgm$ decays presented here is therefore complementary to indirect limits
obtained from searches for rare decays at lower energies.
\section{Summary}
A search for lepton flavour violating decays of the Higgs boson to $\Pe\Pgt$  or $\Pe\Pgm$,
based on the full $\sqrt{s}=8\TeV$ collision data set collected by the CMS experiment in 2012, is presented. No evidence is found for such decays.
Observed upper limits of $\mathcal{B}(\PH \to \Pe \Pgt )<0.69\%$
and $\mathcal{B}(\PH\to\Pe\Pgm )<0.035\%$  at 95\% CL are set for $\textrm{M}_{\PH} = 125\GeV$.
These limits are used to constrain the $Y_{\Pe\Pgt}$ and $Y_{\Pe\Pgm}$ Yukawa couplings as follows:
$\sqrt{\abs{Y_{\Pe\Pgt}}^{2}+\abs{Y_{\Pgt\Pe}}^{2}}<2.4\times 10^{-3}$ and
$\sqrt{\abs{Y_{\Pe\Pgm}}^2 + \abs{Y_{\Pgm\Pe}}^2} < 5.4 \times 10^{-4}$ at 95\% CL.

\clearpage
\begin{acknowledgments}
We congratulate our colleagues in the CERN accelerator departments for the excellent performance of the LHC and thank the technical and administrative staffs at CERN and at other CMS institutes for their contributions to the success of the CMS effort. In addition, we gratefully acknowledge the computing centres and personnel of the Worldwide LHC Computing Grid for delivering so effectively the computing infrastructure essential to our analyses. Finally, we acknowledge the enduring support for the construction and operation of the LHC and the CMS detector provided by the following funding agencies: BMWFW and FWF (Austria); FNRS and FWO (Belgium); CNPq, CAPES, FAPERJ, and FAPESP (Brazil); MES (Bulgaria); CERN; CAS, MoST, and NSFC (China); COLCIENCIAS (Colombia); MSES and CSF (Croatia); RPF (Cyprus); SENESCYT (Ecuador); MoER, ERC IUT and ERDF (Estonia); Academy of Finland, MEC, and HIP (Finland); CEA and CNRS/IN2P3 (France); BMBF, DFG, and HGF (Germany); GSRT (Greece); OTKA and NIH (Hungary); DAE and DST (India); IPM (Iran); SFI (Ireland); INFN (Italy); MSIP and NRF (Republic of Korea); LAS (Lithuania); MOE and UM (Malaysia); BUAP, CINVESTAV, CONACYT, LNS, SEP, and UASLP-FAI (Mexico); MBIE (New Zealand); PAEC (Pakistan); MSHE and NSC (Poland); FCT (Portugal); JINR (Dubna); MON, RosAtom, RAS and RFBR (Russia); MESTD (Serbia); SEIDI and CPAN (Spain); Swiss Funding Agencies (Switzerland); MST (Taipei); ThEPCenter, IPST, STAR and NSTDA (Thailand); TUBITAK and TAEK (Turkey); NASU and SFFR (Ukraine); STFC (United Kingdom); DOE and NSF (USA).

Individuals have received support from the Marie-Curie programme and the European Research Council and EPLANET (European Union); the Leventis Foundation; the A. P. Sloan Foundation; the Alexander von Humboldt Foundation; the Belgian Federal Science Policy Office; the Fonds pour la Formation \`a la Recherche dans l'Industrie et dans l'Agriculture (FRIA-Belgium); the Agentschap voor Innovatie door Wetenschap en Technologie (IWT-Belgium); the Ministry of Education, Youth and Sports (MEYS) of the Czech Republic; the Council of Science and Industrial Research, India; the HOMING PLUS programme of the Foundation for Polish Science, cofinanced from European Union, Regional Development Fund, the Mobility Plus programme of the Ministry of Science and Higher Education, the OPUS programme contract 2014/13/B/ST2/02543 and contract Sonata-bis DEC-2012/07/E/ST2/01406 of the National Science Center (Poland); the Thalis and Aristeia programmes cofinanced by EU-ESF and the Greek NSRF; the National Priorities Research Program by Qatar National Research Fund; the Programa Clar\'in-COFUND del Principado de Asturias; the Rachadapisek Sompot Fund for Postdoctoral Fellowship, Chulalongkorn University and the Chulalongkorn Academic into Its 2nd Century Project Advancement Project (Thailand); and the Welch Foundation, contract C-1845.
\end{acknowledgments}

\bibliography{auto_generated}

\providecommand{\href}[2]{#2}\begingroup\raggedright\begin{thebibliography}{10}%
\makeatletter
\providecommand{\hrefCMSnoop }[0]{\@secondoftwo}%
\makeatother
\providecommand{\doi}{\texttt{doi:}\begingroup \urlstyle{tt}\Url}

\bibitem{Aad:2012tfa}
\hrefCMSnoop {}{{ATLAS Collaboration}, ``{Observation of a new particle in the
  search for the Standard Model Higgs boson with the ATLAS detector at the
  LHC}'',} \textit{ Phys. Lett. B} \textbf{ 716} (2012) 1,
  \href{http://dx.doi.org/10.1016/j.physletb.2012.08.020}{\doi{10.1016/j.physletb.2012.08.020}},
\href{http://www.arXiv.org/abs/1207.7214}{\texttt{arXiv:1207.7214}}.
%%CITATION = ARXIV:1207.7214;%%.

\bibitem{Chatrchyan:2012ufa}
\hrefCMSnoop {}{{CMS Collaboration}, ``{Observation of a new boson at a mass of
  125 GeV with the CMS experiment at the LHC}'',} \textit{ Phys. Lett. B}
  \textbf{ 716} (2012) 30,
  \href{http://dx.doi.org/10.1016/j.physletb.2012.08.021}{\doi{10.1016/j.physletb.2012.08.021}},
\href{http://www.arXiv.org/abs/1207.7235}{\texttt{arXiv:1207.7235}}.
%%CITATION = ARXIV:1207.7235;%%.

\bibitem{Chatrchyan:2013lba}
\hrefCMSnoop {}{{CMS Collaboration}, ``{Observation of a new boson with mass
  near 125 GeV in pp collisions at $\sqrt{s}$ = 7 and 8 TeV}'',} \textit{ JHEP}
  \textbf{ 06} (2013) 081,
  \href{http://dx.doi.org/10.1007/JHEP06(2013)081}{\doi{10.1007/JHEP06(2013)081}},
\href{http://www.arXiv.org/abs/1303.4571}{\texttt{arXiv:1303.4571}}.
%%CITATION = ARXIV:1303.4571;%%.

\bibitem{PhysRevLett.38.622}
\hrefCMSnoop {}{J.~D. Bjorken and S.~Weinberg, ``Mechanism for Nonconservation
  of Muon Number'',} \textit{ Phys. Rev. Lett.} \textbf{ 38} (1977) 622,
  \href{http://dx.doi.org/10.1103/PhysRevLett.38.622}{\doi{10.1103/PhysRevLett.38.622}}.

\bibitem{DiazCruz:1999xe}
\hrefCMSnoop {}{J.~L. Diaz-Cruz and J.~J. Toscano, ``{Lepton flavor violating
  decays of Higgs bosons beyond the standard model}'',} \textit{ Phys. Rev. D}
  \textbf{ 62} (2000) 116005,
  \href{http://dx.doi.org/10.1103/PhysRevD.62.116005}{\doi{10.1103/PhysRevD.62.116005}},
\href{http://www.arXiv.org/abs/hep-ph/9910233}{\texttt{arXiv:hep-ph/9910233}}.
%%CITATION = HEP-PH/9910233;%%.

\bibitem{Han:2000jz}
\hrefCMSnoop {}{T.~Han and D.~Marfatia, ``$h \to \mu \tau$ at Hadron
  Colliders'',} \textit{ Phys. Rev. Lett.} \textbf{ 86} (2001) 1442,
  \href{http://dx.doi.org/10.1103/PhysRevLett.86.1442}{\doi{10.1103/PhysRevLett.86.1442}},
\href{http://www.arXiv.org/abs/hep-ph/0008141}{\texttt{arXiv:hep-ph/0008141}}.
%%CITATION = HEP-PH/0008141;%%.

\bibitem{Arganda:2004bz}
\hrefCMSnoop {}{E.~Arganda, A.~M. Curiel, M.~J. Herrero, and D.~Temes, ``Lepton
  flavor violating {H}iggs boson decays from massive seesaw neutrinos'',}
  \textit{ Phys. Rev. D} \textbf{ 71} (2005) 035011,
  \href{http://dx.doi.org/10.1103/PhysRevD.71.035011}{\doi{10.1103/PhysRevD.71.035011}},
\href{http://www.arXiv.org/abs/hep-ph/0407302}{\texttt{arXiv:hep-ph/0407302}}.
%%CITATION = HEP-PH/0407302;%%.

\bibitem{Arhrib:2012ax}
\hrefCMSnoop {}{A.~Arhrib, Y.~Cheng, and O.~C.~W. Kong, ``{Comprehensive
  analysis on lepton flavor violating Higgs boson to $\mu\bar{\tau} + \tau
  \bar{\mu}$ decay in supersymmetry without R parity}'',} \textit{ Phys. Rev.
  D} \textbf{ 87} (2013) 015025,
  \href{http://dx.doi.org/10.1103/PhysRevD.87.015025}{\doi{10.1103/PhysRevD.87.015025}},
\href{http://www.arXiv.org/abs/1210.8241}{\texttt{arXiv:1210.8241}}.
%%CITATION = ARXIV:1210.8241;%%.

\bibitem{Arana-Catania:2013xma}
\hrefCMSnoop {}{M.~Arana-Catania, E.~Arganda, and M.~J. Herrero,
  ``Non-decoupling {SUSY} in {LFV} {H}iggs decays: a window to new physics at
  the {LHC}'',} \textit{ JHEP} \textbf{ 09} (2013) 160,
  \href{http://dx.doi.org/10.1007/JHEP09(2013)160}{\doi{10.1007/JHEP09(2013)160}},
  \href{http://www.arXiv.org/abs/1304.3371}{\texttt{arXiv:1304.3371}}.
[Erratum: \DOI{10.1007/JHEP10(2015)192}].
%%CITATION = ARXIV:1304.3371;%%.

\bibitem{Arganda:2015uca}
\hrefCMSnoop {}{E.~Arganda, M.~J. Herrero, R.~Morales, and A.~Szynkman,
  ``Analysis of the {$h, H, A \to\tau\mu$} decays induced from {SUSY} loops
  within the {M}ass {I}nsertion {A}pproximation'',} \textit{ JHEP} \textbf{ 03}
  (2016) 055,
  \href{http://dx.doi.org/10.1007/JHEP03(2016)055}{\doi{10.1007/JHEP03(2016)055}},
\href{http://www.arXiv.org/abs/1510.04685}{\texttt{arXiv:1510.04685}}.
%%CITATION = ARXIV:1510.04685;%%.

\bibitem{Arganda:2015naa}
\hrefCMSnoop {}{E.~Arganda, M.~J. Herrero, X.~Marcano, and C.~Weiland,
  ``Enhancement of the lepton flavor violating {H}iggs boson decay rates from
  {SUSY} loops in the inverse seesaw model'',} \textit{ Phys. Rev. D} \textbf{
  93} (2016), no.~5, 055010,
  \href{http://dx.doi.org/10.1103/PhysRevD.93.055010}{\doi{10.1103/PhysRevD.93.055010}},
\href{http://www.arXiv.org/abs/1508.04623}{\texttt{arXiv:1508.04623}}.
%%CITATION = ARXIV:1508.04623;%%.

\bibitem{Agashe:2009di}
\hrefCMSnoop {}{K.~Agashe and R.~Contino, ``{Composite Higgs-mediated
  flavor-changing neutral current}'',} \textit{ Phys. Rev. D} \textbf{ 80}
  (2009) 075016,
  \href{http://dx.doi.org/10.1103/PhysRevD.80.075016}{\doi{10.1103/PhysRevD.80.075016}},
\href{http://www.arXiv.org/abs/0906.1542}{\texttt{arXiv:0906.1542}}.
%%CITATION = ARXIV:0906.1542;%%.

\bibitem{Azatov:2009na}
\hrefCMSnoop {}{A.~Azatov, M.~Toharia, and L.~Zhu, ``{Higgs mediated flavor
  changing neutral currents in warped extra dimensions}'',} \textit{ Phys. Rev.
  D} \textbf{ 80} (2009) 035016,
  \href{http://dx.doi.org/10.1103/PhysRevD.80.035016}{\doi{10.1103/PhysRevD.80.035016}},
\href{http://www.arXiv.org/abs/0906.1990}{\texttt{arXiv:0906.1990}}.
%%CITATION = ARXIV:0906.1990;%%.

\bibitem{Ishimori:2010au}
H.~Ishimori\hrefCMSnoop {}{ {et~al.}, ``Non-{A}belian Discrete Symmetries in
  Particle Physics'',} \textit{ {Prog. Theor. Phys. Suppl.}} \textbf{ 183}
  (2010) 1,
  \href{http://dx.doi.org/10.1143/PTPS.183.1}{\doi{10.1143/PTPS.183.1}},
\href{http://www.arXiv.org/abs/1003.3552}{\texttt{arXiv:1003.3552}}.
%%CITATION = ARXIV:1003.3552;%%.

\bibitem{Perez:2008ee}
\hrefCMSnoop {}{G.~Perez and L.~Randall, ``{Natural neutrino masses and mixings
  from warped geometry}'',} \textit{ JHEP} \textbf{ 01} (2009) 077,
  \href{http://dx.doi.org/10.1088/1126-6708/2009/01/077}{\doi{10.1088/1126-6708/2009/01/077}},
\href{http://www.arXiv.org/abs/0805.4652}{\texttt{arXiv:0805.4652}}.
%%CITATION = ARXIV:0805.4652;%%.

\bibitem{Casagrande:2008hr}
S.~Casagrande\hrefCMSnoop {}{ {et~al.}, ``{Flavor physics in the
  Randall-Sundrum model I. Theoretical setup and electroweak precision
  tests}'',} \textit{ JHEP} \textbf{ 10} (2008) 094,
  \href{http://dx.doi.org/10.1088/1126-6708/2008/10/094}{\doi{10.1088/1126-6708/2008/10/094}},
\href{http://www.arXiv.org/abs/0807.4937}{\texttt{arXiv:0807.4937}}.
%%CITATION = ARXIV:0807.4937;%%.

\bibitem{Buras:2009ka}
\hrefCMSnoop {}{A.~J. Buras, B.~Duling, and S.~Gori, ``{The impact of
  Kaluza-Klein fermions on Standard Model fermion couplings in a RS model with
  custodial protection}'',} \textit{ JHEP} \textbf{ 09} (2009) 076,
  \href{http://dx.doi.org/10.1088/1126-6708/2009/09/076}{\doi{10.1088/1126-6708/2009/09/076}},
\href{http://www.arXiv.org/abs/0905.2318}{\texttt{arXiv:0905.2318}}.
%%CITATION = ARXIV:0905.2318;%%.

\bibitem{Blanke:2008zb}
M.~Blanke\hrefCMSnoop {}{ {et~al.}, ``{$\Delta F=2$ observables and fine-tuning
  in a warped extra dimension with custodial protection }'',} \textit{ JHEP}
  \textbf{ 03} (2009) 001,
  \href{http://dx.doi.org/10.1088/1126-6708/2009/03/001}{\doi{10.1088/1126-6708/2009/03/001}},
\href{http://www.arXiv.org/abs/0809.1073}{\texttt{arXiv:0809.1073}}.
%%CITATION = ARXIV:0809.1073;%%.

\bibitem{Giudice:2008uua}
\hrefCMSnoop {}{G.~F. Giudice and O.~Lebedev, ``{Higgs-dependent Yukawa
  couplings}'',} \textit{ Phys. Lett. B} \textbf{ 665} (2008) 79,
  \href{http://dx.doi.org/10.1016/j.physletb.2008.05.062}{\doi{10.1016/j.physletb.2008.05.062}},
\href{http://www.arXiv.org/abs/0804.1753}{\texttt{arXiv:0804.1753}}.
%%CITATION = ARXIV:0804.1753;%%.

\bibitem{AguilarSaavedra:2009mx}
\hrefCMSnoop {}{J.~A. Aguilar-Saavedra, ``{A minimal set of top-Higgs anomalous
  couplings}'',} \textit{ Nucl. Phys. B} \textbf{ 821} (2009) 215,
  \href{http://dx.doi.org/10.1016/j.nuclphysb.2009.06.022}{\doi{10.1016/j.nuclphysb.2009.06.022}},
\href{http://www.arXiv.org/abs/0904.2387}{\texttt{arXiv:0904.2387}}.
%%CITATION = ARXIV:0904.2387;%%.

\bibitem{Albrecht:2009xr}
M.~E. Albrecht\hrefCMSnoop {}{ {et~al.}, ``{Electroweak and flavour structure
  of a warped extra dimension with custodial protection}'',} \textit{ JHEP}
  \textbf{ 09} (2009) 064,
  \href{http://dx.doi.org/10.1088/1126-6708/2009/09/064}{\doi{10.1088/1126-6708/2009/09/064}},
\href{http://www.arXiv.org/abs/0903.2415}{\texttt{arXiv:0903.2415}}.
%%CITATION = ARXIV:0903.2415;%%.

\bibitem{Goudelis:2011un}
\hrefCMSnoop {}{A.~Goudelis, O.~Lebedev, and J.~H. Park, ``{Higgs-induced
  lepton flavor violation}'',} \textit{ Phys. Lett. B} \textbf{ 707} (2012)
  369,
  \href{http://dx.doi.org/10.1016/j.physletb.2011.12.059}{\doi{10.1016/j.physletb.2011.12.059}},
\href{http://www.arXiv.org/abs/1111.1715}{\texttt{arXiv:1111.1715}}.
%%CITATION = ARXIV:1111.1715;%%.

\bibitem{McKeen:2012av}
\hrefCMSnoop {}{D.~McKeen, M.~Pospelov, and A.~Ritz, ``{Modified Higgs
  branching ratios versus CP and lepton flavor violation}'',} \textit{ Phys.
  Rev. D} \textbf{ 86} (2012) 113004,
  \href{http://dx.doi.org/10.1103/PhysRevD.86.113004}{\doi{10.1103/PhysRevD.86.113004}},
\href{http://www.arXiv.org/abs/1208.4597}{\texttt{arXiv:1208.4597}}.
%%CITATION = ARXIV:1208.4597;%%.

\bibitem{Pilaftsis199268}
\hrefCMSnoop {}{A.~Pilaftsis, ``Lepton flavour nonconservation in {$\PH^0$}
  decays'',} \textit{ Phys. Lett. B} \textbf{ 285} (1992) 68,
  \href{http://dx.doi.org/10.1016/0370-2693(92)91301-O}{\doi{10.1016/0370-2693(92)91301-O}}.

\bibitem{PhysRevD.47.1080}
\hrefCMSnoop {}{J.~G. K{\"o}rner, A.~Pilaftsis, and K.~Schilcher, ``Leptonic
  $\mathrm{CP}$ asymmetries in flavor-changing {$\PH^{0}$} decays'',} \textit{
  Phys. Rev. D} \textbf{ 47} (1993) 1080,
  \href{http://dx.doi.org/10.1103/PhysRevD.47.1080}{\doi{10.1103/PhysRevD.47.1080}}.

\bibitem{Arganda:2014dta}
\hrefCMSnoop {}{E.~Arganda, M.~J. Herrero, X.~Marcano, and C.~Weiland,
  ``Imprints of massive inverse seesaw model neutrinos in lepton flavor
  violating {H}iggs boson decays'',} \textit{ Phys. Rev. D} \textbf{ 91}
  (2015), no.~1, 015001,
  \href{http://dx.doi.org/10.1103/PhysRevD.91.015001}{\doi{10.1103/PhysRevD.91.015001}},
\href{http://www.arXiv.org/abs/1405.4300}{\texttt{arXiv:1405.4300}}.
%%CITATION = ARXIV:1405.4300;%%.

\bibitem{Khachatryan:2015kon}
\hrefCMSnoop {}{{CMS Collaboration}, ``{Search for lepton-flavour-violating
  decays of the Higgs boson}'',} \textit{ Phys. Lett. B} \textbf{ 749} (2015)
  337,
  \href{http://dx.doi.org/10.1016/j.physletb.2015.07.053}{\doi{10.1016/j.physletb.2015.07.053}},
\href{http://www.arXiv.org/abs/1502.07400}{\texttt{arXiv:1502.07400}}.
%%CITATION = ARXIV:1502.07400;%%.

\bibitem{Aad:2015gha}
\hrefCMSnoop {}{{ATLAS Collaboration}, ``{Search for lepton-flavour-violating
  $\PH\to\mu\tau$ decays of the Higgs boson with the ATLAS detector}'',}
  \textit{ JHEP} \textbf{ 11} (2015) 211,
  \href{http://dx.doi.org/10.1007/JHEP11(2015)211}{\doi{10.1007/JHEP11(2015)211}},
  \href{http://www.arXiv.org/abs/1508.03372}{\texttt{arXiv:1508.03372}}.

\bibitem{Aad:2016blu}
\hrefCMSnoop {}{{ATLAS Collaboration}, ``{Search for lepton-flavour-violating
  decays of the Higgs and $Z$ bosons with the ATLAS detector}'',} (2016).
  \href{http://www.arXiv.org/abs/1604.07730}{\texttt{arXiv:1604.07730}}.
Submitted to {JHEP}.
%%CITATION = ARXIV:1604.07730;%%.

\bibitem{McWilliams:1980kj}
\hrefCMSnoop {}{B.~McWilliams and L.-F. Li, ``{Virtual effects of Higgs
  particles}'',} \textit{ Nucl. Phys. B} \textbf{ 179} (1981) 62,
\href{http://dx.doi.org/10.1016/0550-3213(81)90249-2}{\doi{10.1016/0550-3213(81)90249-2}}.
%%CITATION = NUPHA,B179,62;%%.

\bibitem{Shanker:1981mj}
\hrefCMSnoop {}{O.~U. Shanker, ``{Flavour violation, scalar particles and
  leptoquarks}'',} \textit{ Nucl. Phys. B} \textbf{ 206} (1982) 253,
\href{http://dx.doi.org/10.1016/0550-3213(82)90534-X}{\doi{10.1016/0550-3213(82)90534-X}}.
%%CITATION = NUPHA,B206,253;%%.

\bibitem{Blankenburg:2012ex}
\hrefCMSnoop {}{G.~Blankenburg, J.~Ellis, and G.~Isidori, ``{Flavour-changing
  decays of a 125 GeV Higgs-like particle}'',} \textit{ Phys. Lett. B} \textbf{
  712} (2012) 386,
  \href{http://dx.doi.org/10.1016/j.physletb.2012.05.007}{\doi{10.1016/j.physletb.2012.05.007}},
\href{http://www.arXiv.org/abs/1202.5704}{\texttt{arXiv:1202.5704}}.
%%CITATION = ARXIV:1202.5704;%%.

\bibitem{Harnik:2012pb}
\hrefCMSnoop {}{R.~Harnik, J.~Kopp, and J.~Zupan, ``Flavor violating Higgs
  decays'',} \textit{ JHEP} \textbf{ 03} (2013) 26,
  \href{http://dx.doi.org/10.1007/JHEP03(2013)026}{\doi{10.1007/JHEP03(2013)026}},
  \href{http://www.arXiv.org/abs/1209.1397}{\texttt{arXiv:1209.1397}}.

\bibitem{Beringer:1900zz}
\hrefCMSnoop {}{{Particle Data Group}, J.~Beringer {et~al.}, ``{Review of
  Particle Physics}'',} \textit{ Phys. Rev. D} \textbf{ 86} (2012) 010001,
\href{http://dx.doi.org/10.1103/PhysRevD.86.010001}{\doi{10.1103/PhysRevD.86.010001}}.
%%CITATION = PHRVA,D86,010001;%%.

\bibitem{Celis:2013xja}
\hrefCMSnoop {}{A.~Celis, V.~Cirigliano, and E.~Passemar, ``{Lepton flavor
  violation in the Higgs sector and the role of hadronic tau-lepton decays}'',}
  \textit{ Phys. Rev. D} \textbf{ 89} (2014) 013008,
  \href{http://dx.doi.org/10.1103/PhysRevD.89.013008}{\doi{10.1103/PhysRevD.89.013008}},
\href{http://www.arXiv.org/abs/1309.3564}{\texttt{arXiv:1309.3564}}.
%%CITATION = ARXIV:1309.3564;%%.

\bibitem{Barr:1990vd}
\hrefCMSnoop {}{S.~M. Barr and A.~Zee, ``Electric Dipole Moment of the Electron
  and of the Neutron'',} \textit{ Phys. Rev. Lett.} \textbf{ 65} (1990) 21,
\href{http://dx.doi.org/10.1103/PhysRevLett.65.21}{\doi{10.1103/PhysRevLett.65.21}}.
%%CITATION = PRLTA,65,21;%%.

\bibitem{Chatrchyan:2014vua}
\hrefCMSnoop {}{{CMS Collaboration}, ``{Evidence for the direct decay of the
  125 GeV Higgs boson to fermions}'',} \textit{ Nature Phys.} \textbf{ 10}
  (2014) 557,
  \href{http://dx.doi.org/10.1038/nphys3005}{\doi{10.1038/nphys3005}},
\href{http://www.arXiv.org/abs/1401.6527}{\texttt{arXiv:1401.6527}}.
%%CITATION = ARXIV:1401.6527;%%.

\bibitem{CMS-PAPERS-HIG-13-004}
\hrefCMSnoop {}{{CMS Collaboration}, ``Evidence for the 125\GeV Higgs boson
  decaying to a pair of $\tau$ leptons'',} \textit{ JHEP} \textbf{ 05} (2014)
  104,
  \href{http://dx.doi.org/10.1007/JHEP05(2014)104}{\doi{10.1007/JHEP05(2014)104}},
  \href{http://www.arXiv.org/abs/1401.5041}{\texttt{arXiv:1401.5041}}.

\bibitem{Aad:2015vsa}
\hrefCMSnoop {}{{ATLAS Collaboration}, ``{Evidence for the Higgs-boson Yukawa
  coupling to tau leptons with the ATLAS detector}'',} \textit{ JHEP} \textbf{
  04} (2015) 117,
  \href{http://dx.doi.org/10.1007/JHEP04(2015)117}{\doi{10.1007/JHEP04(2015)117}},
\href{http://www.arXiv.org/abs/1501.04943}{\texttt{arXiv:1501.04943}}.
%%CITATION = ARXIV:1501.04943;%%.

\bibitem{CMS-JINST}
\hrefCMSnoop {}{{CMS Collaboration}, ``The {CMS} experiment at the {CERN}
  {LHC}'',} \textit{ JINST} \textbf{ 3} (2008) S08004,
\href{http://dx.doi.org/10.1088/1748-0221/3/08/S08004}{\doi{10.1088/1748-0221/3/08/S08004}}.
%%CITATION = JINST,3,S08004;%%.

\bibitem{GEANT4}
\hrefCMSnoop {}{{GEANT4} Collaboration, ``{GEANT4} --- a simulation toolkit'',}
  \textit{ Nucl. Instrum. Meth. A} \textbf{ 506} (2003) 250,
\href{http://dx.doi.org/10.1016/S0168-9002(03)01368-8}{\doi{10.1016/S0168-9002(03)01368-8}}.
%%CITATION = NUIMA,A506,250;%%.

\bibitem{Georgi:1977gs}
\hrefCMSnoop {}{H.~M. Georgi, S.~L. Glashow, M.~E. Machacek, and D.~V.
  Nanopoulos, ``{H}iggs Bosons from Two Gluon Annihilation in Proton Proton
  Collisions'',} \textit{ Phys. Rev. Lett.} \textbf{ 40} (1978) 692,
\href{http://dx.doi.org/10.1103/PhysRevLett.40.692}{\doi{10.1103/PhysRevLett.40.692}}.
%%CITATION = PRLTA,40,692;%%.

\bibitem{Cahn:1986zv}
\hrefCMSnoop {}{R.~N. Cahn, S.~D. Ellis, R.~Kleiss, and W.~J. Stirling,
  ``{Transverse-momentum signatures for heavy Higgs bosons}'',} \textit{ Phys.
  Rev. D} \textbf{ 35} (1987) 1626,
\href{http://dx.doi.org/10.1103/PhysRevD.35.1626}{\doi{10.1103/PhysRevD.35.1626}}.
%%CITATION = PHRVA,D35,1626;%%.

\bibitem{Glashow:1978ab}
\hrefCMSnoop {}{S.~L. Glashow, D.~V. Nanopoulos, and A.~Yildiz, ``{Associated
  production of Higgs bosons and Z particles}'',} \textit{ Phys. Rev. D}
  \textbf{ 18} (1978) 1724,
\href{http://dx.doi.org/10.1103/PhysRevD.18.1724}{\doi{10.1103/PhysRevD.18.1724}}.
%%CITATION = PHRVA,D18,1724;%%.

\bibitem{Sjostrand:pythia8}
\hrefCMSnoop {}{T.~Sj{\"o}strand, S.~Mrenna, and P.~Skands, ``A brief
  introduction to {PYTHIA} 8.1'',} \textit{ {Comput. Phys. Commun.}} \textbf{
  178} (2007) 852,
  \href{http://dx.doi.org/10.1016/j.cpc.2008.01.036}{\doi{10.1016/j.cpc.2008.01.036}},
\href{http://www.arXiv.org/abs/0710.3820}{\texttt{arXiv:0710.3820}}.
%%CITATION = HEP-PH/0603175;%%.

\bibitem{pythia}
\hrefCMSnoop {}{T.~Sj{\"o}strand, S.~Mrenna, and P.~Skands, ``{PYTHIA} 6.4
  physics and manual'',} \textit{ JHEP} \textbf{ 05} (2006) 026,
  \href{http://dx.doi.org/10.1088/1126-6708/2006/05/026}{\doi{10.1088/1126-6708/2006/05/026}},
\href{http://www.arXiv.org/abs/hep-ph/0603175}{\texttt{arXiv:hep-ph/0603175}}.
%%CITATION = HEP-PH/0603175;%%.

\bibitem{Nadolsky:2008zw}
P.~M. Nadolsky\hrefCMSnoop {}{ {et~al.}, ``{Implications of CTEQ global
  analysis for collider observables}'',} \textit{ Phys. Rev. D} \textbf{ 78}
  (2008) 013004,
  \href{http://dx.doi.org/10.1103/PhysRevD.78.013004}{\doi{10.1103/PhysRevD.78.013004}},
\href{http://www.arXiv.org/abs/0802.0007}{\texttt{arXiv:0802.0007}}.
%%CITATION = ARXIV:0802.0007;%%.

\bibitem{Nason:2004rx}
\hrefCMSnoop {}{P.~Nason, ``{A new method for combining NLO QCD with shower
  Monte Carlo algorithms}'',} \textit{ JHEP} \textbf{ 11} (2004) 040,
  \href{http://dx.doi.org/10.1088/1126-6708/2004/11/040}{\doi{10.1088/1126-6708/2004/11/040}},
\href{http://www.arXiv.org/abs/hep-ph/0409146}{\texttt{arXiv:hep-ph/0409146}}.
%%CITATION = HEP-PH/0409146;%%.

\bibitem{Frixione:2007vw}
\hrefCMSnoop {}{S.~Frixione, P.~Nason, and C.~Oleari, ``{Matching NLO QCD
  computations with parton shower simulations: the POWHEG method}'',} \textit{
  JHEP} \textbf{ 11} (2007) 070,
  \href{http://dx.doi.org/10.1088/1126-6708/2007/11/070}{\doi{10.1088/1126-6708/2007/11/070}},
\href{http://www.arXiv.org/abs/0709.2092}{\texttt{arXiv:0709.2092}}.
%%CITATION = ARXIV:0709.2092;%%.

\bibitem{Alioli:2010xd}
\hrefCMSnoop {}{S.~Alioli, P.~Nason, C.~Oleari, and E.~Re, ``{A general
  framework for implementing NLO calculations in shower Monte Carlo programs:
  the POWHEG BOX}'',} \textit{ JHEP} \textbf{ 06} (2010) 043,
  \href{http://dx.doi.org/10.1007/JHEP06(2010)043}{\doi{10.1007/JHEP06(2010)043}},
\href{http://www.arXiv.org/abs/1002.2581}{\texttt{arXiv:1002.2581}}.
%%CITATION = ARXIV:1002.2581;%%.

\bibitem{Alioli:2010xa}
S.~Alioli\hrefCMSnoop {}{ {et~al.}, ``{Jet pair production in POWHEG}'',}
  \textit{ JHEP} \textbf{ 04} (2011) 081,
  \href{http://dx.doi.org/10.1007/JHEP04(2011)081}{\doi{10.1007/JHEP04(2011)081}},
\href{http://www.arXiv.org/abs/1012.3380}{\texttt{arXiv:1012.3380}}.
%%CITATION = ARXIV:1012.3380;%%.

\bibitem{Alioli:2008tz}
\hrefCMSnoop {}{S.~Alioli, P.~Nason, C.~Oleari, and E.~Re, ``{NLO Higgs boson
  production via gluon fusion matched with shower in POWHEG}'',} \textit{ JHEP}
  \textbf{ 04} (2009) 002,
  \href{http://dx.doi.org/10.1088/1126-6708/2009/04/002}{\doi{10.1088/1126-6708/2009/04/002}},
\href{http://www.arXiv.org/abs/0812.0578}{\texttt{arXiv:0812.0578}}.
%%CITATION = ARXIV:0812.0578;%%.

\bibitem{Alwall:2011uj}
J.~Alwall\hrefCMSnoop {}{ {et~al.}, ``{MadGraph 5: going beyond}'',} \textit{
  JHEP} \textbf{ 06} (2011) 128,
  \href{http://dx.doi.org/10.1007/JHEP06(2011)128}{\doi{10.1007/JHEP06(2011)128}},
\href{http://www.arXiv.org/abs/1106.0522}{\texttt{arXiv:1106.0522}}.
%%CITATION = ARXIV:1106.0522;%%.

\bibitem{Field:2010bc}
\href {http://inspirehep.net/record/873443/files/arXiv:1010.3558.pdf}{R.~Field,
  ``{Early LHC Underlying Event Data - Findings and Surprises}'',} in \textit{
  {Hadron collider physics. Proceedings, 22nd Conference, HCP 2010, Toronto,
  Canada, August 23-27, 2010}}.
\newblock 2010.
\newblock
\href{http://www.arXiv.org/abs/1010.3558}{\texttt{arXiv:1010.3558}}.
\newblock
%%CITATION = ARXIV:1010.3558;%%.

\bibitem{Chatrchyan:2014fea}
\hrefCMSnoop {}{{CMS Collaboration}, ``{Description and performance of track
  and primary-vertex reconstruction with the CMS tracker}'',} \textit{ JINST}
  \textbf{ 9} (2014) P10009,
  \href{http://dx.doi.org/10.1088/1748-0221/9/10/P10009}{\doi{10.1088/1748-0221/9/10/P10009}},
\href{http://www.arXiv.org/abs/1405.6569}{\texttt{arXiv:1405.6569}}.
%%CITATION = ARXIV:1405.6569;%%.

\bibitem{CMS-PAS-PFT-09-001}
\href {http://cdsweb.cern.ch/record/1194487}{{CMS Collaboration},
  ``Particle--Flow Event Reconstruction in {CMS} and Performance for Jets,
  Taus, and {\MET}'',} CMS Physics Analysis Summary CMS-PAS-PFT-09-001, 2009.

\bibitem{CMS-PAS-PFT-10-002}
\href {http://cds.cern.ch/record/1247373}{{CMS Collaboration}, ``Particle flow
  reconstruction of 0.9 TeV and 2.36 TeV collision events in CMS'',} CMS
  Physics Analysis Note CMS-PAS-PFT-10-001, 2010.

\bibitem{CMS-PAS-PFT-10-003}
\href {http://cdsweb.cern.ch/record/1279347}{{CMS Collaboration},
  ``Commissioning of the particle--flow event reconstruction with leptons from
  J/$\psi$ and $\PW$ decays at 7 TeV'',} CMS Physics Analysis Summary
  CMS-PAS-PFT-10-003, 2010.

\bibitem{Khachatryan:2014gga}
\hrefCMSnoop {}{{CMS Collaboration}, ``{Performance of the CMS missing
  transverse momentum reconstruction in pp data at $\sqrt{s}$ = 8 TeV}'',}
  \textit{ JINST} \textbf{ 10} (2015) P02006,
  \href{http://dx.doi.org/10.1088/1748-0221/10/02/P02006}{\doi{10.1088/1748-0221/10/02/P02006}},
\href{http://www.arXiv.org/abs/1411.0511}{\texttt{arXiv:1411.0511}}.
%%CITATION = ARXIV:1411.0511;%%.

\bibitem{Khachatryan:2015hwa}
\hrefCMSnoop {}{{CMS Collaboration}, ``{Performance of electron reconstruction
  and selection with the CMS detector in proton-proton collisions at $\sqrt{s}$
  = 8 TeV}'',} \textit{ JINST} \textbf{ 10} (2015) P06005,
  \href{http://dx.doi.org/10.1088/1748-0221/10/06/P06005}{\doi{10.1088/1748-0221/10/06/P06005}},
\href{http://www.arXiv.org/abs/1502.02701}{\texttt{arXiv:1502.02701}}.
%%CITATION = ARXIV:1502.02701;%%.

\bibitem{HIG-13-002}
\hrefCMSnoop {}{{CMS Collaboration}, ``Measurement of the properties of a Higgs
  boson in the four-lepton final state'',} \textit{ Phys. Rev. D} \textbf{
  {89}} (2014) 092007,
  \href{http://dx.doi.org/10.1103/PhysRevD.89.092007}{\doi{10.1103/PhysRevD.89.092007}},
  \href{http://www.arXiv.org/abs/1312.5353}{\texttt{arXiv:1312.5353}}.

\bibitem{ref:Muscle}
\hrefCMSnoop {}{{CMS Collaboration}, ``{Performance of CMS muon reconstruction
  in pp collision events at $\sqrt{s} = 7$ TeV}'',} \textit{ JINST} \textbf{ 7}
  (2012) P10002,
  \href{http://dx.doi.org/10.1088/1748-0221/7/10/P10002}{\doi{10.1088/1748-0221/7/10/P10002}},
\href{http://www.arXiv.org/abs/1206.4071}{\texttt{arXiv:1206.4071}}.
%%CITATION = ARXIV:1206.4071;%%.

\bibitem{Khachatryan:2015dfa}
\hrefCMSnoop {}{{CMS Collaboration}, ``{Reconstruction and identification of
  tau lepton decays to hadrons and $\nu_\tau$ at CMS}'',} \textit{ JINST}
  \textbf{ 11} (2016) P01019,
  \href{http://dx.doi.org/10.1088/1748-0221/11/01/P01019}{\doi{10.1088/1748-0221/11/01/P01019}},
\href{http://www.arXiv.org/abs/1510.07488}{\texttt{arXiv:1510.07488}}.
%%CITATION = ARXIV:1510.07488;%%.

\bibitem{Cacciari:2011ma}
\hrefCMSnoop {}{M.~Cacciari, G.~P. Salam, and G.~Soyez, ``{FastJet user
  manual}'',} \textit{ Eur. Phys. J. C} \textbf{ 72} (2012) 1896,
  \href{http://dx.doi.org/10.1140/epjc/s10052-012-1896-2}{\doi{10.1140/epjc/s10052-012-1896-2}},
\href{http://www.arXiv.org/abs/1111.6097}{\texttt{arXiv:1111.6097}}.
%%CITATION = ARXIV:1111.6097;%%.

\bibitem{Cacciari:2008gp}
\hrefCMSnoop {}{M.~Cacciari, G.~P. Salam, and G.~Soyez, ``The anti-$k_t$ jet
  clustering algorithm'',} \textit{ JHEP} \textbf{ 04} (2008) 063,
  \href{http://dx.doi.org/10.1088/1126-6708/2008/04/063}{\doi{10.1088/1126-6708/2008/04/063}},
  \href{http://www.arXiv.org/abs/0802.1189}{\texttt{arXiv:0802.1189}}.

\bibitem{CMS-JME-10-011}
\hrefCMSnoop {}{{CMS Collaboration}, ``Determination of jet energy calibration
  and transverse momentum resolution in {CMS}'',} \textit{ JINST} \textbf{ 6}
  (2011) 11002,
  \href{http://dx.doi.org/10.1088/1748-0221/6/11/P11002}{\doi{10.1088/1748-0221/6/11/P11002}},
  \href{http://www.arXiv.org/abs/1107.4277}{\texttt{arXiv:1107.4277}}.

\bibitem{CMS-PAS-JME-13-005}
\href {http://cdsweb.cern.ch/record/1581583}{{CMS Collaboration}, ``Pile-up Jet
  Identification'',} CMS Physics Analysis Summary CMS-PAS-JME-13-005, 2013.

\bibitem{Ellis:1987xu}
\hrefCMSnoop {}{R.~K. Ellis, I.~Hinchliffe, M.~Soldate, and J.~J. van~der Bij,
  ``{Higgs Decay to $\tau^+\tau^-$: A possible signature of intermediate mass
  Higgs bosons at high energy hadron colliders}'',} \textit{ Nucl. Phys. B}
  \textbf{ 297} (1988) 221,
\href{http://dx.doi.org/10.1016/0550-3213(88)90019-3}{\doi{10.1016/0550-3213(88)90019-3}}.
%%CITATION = NUPHA,B297,221;%%.

\bibitem{CMS-PAS-BTV-13-001}
\href {https://cds.cern.ch/record/1581306}{{CMS Collaboration}, ``{Performance
  of b tagging at $\sqrt{s}$ = 8 TeV in multijet, ttbar and boosted topology
  events}'',} CMS Physics Analysis Summary CMS-PAS-BTV-13-001, 2013.

\bibitem{Junk}
\hrefCMSnoop {}{T.~Junk, ``Confidence level computation for combining searches
  with small statistics'',} \textit{ Nucl. Instrum. Meth. A} \textbf{ 434}
  (1999) 435,
  \href{http://dx.doi.org/10.1016/S0168-9002(99)00498-2}{\doi{10.1016/S0168-9002(99)00498-2}},
  \href{http://www.arXiv.org/abs/hep-ex/9902006}{\texttt{arXiv:hep-ex/9902006}}.

\bibitem{Read2}
\hrefCMSnoop {}{A.~L. Read, ``Presentation of search results: the {$CL_s$}
  technique'',} \textit{ J. Phys. G} \textbf{ 28} (2002) 2693,
  \href{http://dx.doi.org/10.1088/0954-3899/28/10/313}{\doi{10.1088/0954-3899/28/10/313}}.

\bibitem{CMS:2011aa}
\hrefCMSnoop {}{{CMS Collaboration}, ``{Measurement of the inclusive $\PW$ and
  $\cPZ$ production cross sections in pp collisions at $\sqrt{s}=7$ TeV with
  the CMS experiment}'',} \textit{ JHEP} \textbf{ 10} (2011) 132,
  \href{http://dx.doi.org/10.1007/JHEP10(2011)132}{\doi{10.1007/JHEP10(2011)132}},
\href{http://www.arXiv.org/abs/1107.4789}{\texttt{arXiv:1107.4789}}.
%%CITATION = ARXIV:1107.4789;%%.

\bibitem{Chatrchyan:2014mua}
\hrefCMSnoop {}{{CMS Collaboration}, ``Measurement of Inclusive {$W$} and {$Z$}
  Boson Production Cross Sections in $pp$ Collisions at {$\sqrt{s}$ = 8
  TeV}'',} \textit{ Phys. Rev. Lett.} \textbf{ 112} (2014) 191802,
  \href{http://dx.doi.org/10.1103/PhysRevLett.112.191802}{\doi{10.1103/PhysRevLett.112.191802}},
\href{http://www.arXiv.org/abs/1402.0923}{\texttt{arXiv:1402.0923}}.
%%CITATION = ARXIV:1402.0923;%%.

\bibitem{Chatrchyan:2012nj}
\hrefCMSnoop {}{{CMS Collaboration}, ``Measurement of the inelastic
  proton-proton cross section at {$\sqrt{s} = 7\TeV$}'',} \textit{ Phys. Lett.
  B} \textbf{ 722} (2013) 5,
  \href{http://dx.doi.org/10.1016/j.physletb.2013.03.024}{\doi{10.1016/j.physletb.2013.03.024}}.

\bibitem{Chatrchyan2013190}
\hrefCMSnoop {}{{CMS Collaboration}, ``Measurement of the {$\PW^+\PW^-$} and
  {$\cPZ\cPZ$} production cross sections in pp collisions at
  {$\sqrt{s}=8$\TeV}'',} \textit{ Phys. Lett. B} \textbf{ 721} (2013) 190,
  \href{http://dx.doi.org/10.1016/j.physletb.2013.03.027}{\doi{10.1016/j.physletb.2013.03.027}},
  \href{http://www.arXiv.org/abs/1301.4698}{\texttt{arXiv:1301.4698}}.

\bibitem{Chatrchyan:2014tua}
\hrefCMSnoop {}{{CMS Collaboration}, ``Observation of the Associated Production
  of a Single Top Quark and a {$W$} Boson in $pp$ Collisions at {$\sqrt s =8$
  TeV}'',} \textit{ Phys. Rev. Lett.} \textbf{ 112} (2014), no.~23, 231802,
  \href{http://dx.doi.org/10.1103/PhysRevLett.112.231802}{\doi{10.1103/PhysRevLett.112.231802}},
\href{http://www.arXiv.org/abs/1401.2942}{\texttt{arXiv:1401.2942}}.
%%CITATION = ARXIV:1401.2942;%%.

\bibitem{Khachatryan:2014loa}
\hrefCMSnoop {}{{CMS Collaboration}, ``{Measurement of the $\mathrm{t \bar t}$
  production cross section in pp collisions at $\sqrt s = 8$ TeV in dilepton
  final states containing one $\tau$ lepton}'',} \textit{ Phys. Lett. B}
  \textbf{ 739} (2014) 23,
  \href{http://dx.doi.org/10.1016/j.physletb.2014.10.032}{\doi{10.1016/j.physletb.2014.10.032}},
\href{http://www.arXiv.org/abs/1407.6643}{\texttt{arXiv:1407.6643}}.
%%CITATION = ARXIV:1407.6643;%%.

\bibitem{Martin:2009iq}
\hrefCMSnoop {}{A.~D. Martin, W.~J. Stirling, R.~S. Thorne, and G.~Watt,
  ``Parton distributions for the {LHC}'',} \textit{ Eur. Phys. J. C} \textbf{
  63} (2009) 189,
  \href{http://dx.doi.org/10.1140/epjc/s10052-009-1072-5}{\doi{10.1140/epjc/s10052-009-1072-5}},
\href{http://www.arXiv.org/abs/0901.0002}{\texttt{arXiv:0901.0002}}.
%%CITATION = ARXIV:0901.0002;%%.

\bibitem{Ball:2010de}
\hrefCMSnoop {}{{NNPDF} Collaboration, ``A first unbiased global {NLO}
  determination of parton distributions and their uncertainties'',} \textit{
  Nucl. Phys. B} \textbf{ 838} (2010) 136,
  \href{http://dx.doi.org/10.1016/j.nuclphysb.2010.05.008}{\doi{10.1016/j.nuclphysb.2010.05.008}},
\href{http://www.arXiv.org/abs/1002.4407}{\texttt{arXiv:1002.4407}}.
%%CITATION = ARXIV:1002.4407;%%.

\bibitem{Botje:2011sn}
M.~Botje\hrefCMSnoop {}{ {et~al.}, ``{The PDF4LHC Working Group Interim
  Recommendations}'',} (2011).
\href{http://www.arXiv.org/abs/1101.0538}{\texttt{arXiv:1101.0538}}.
%%CITATION = ARXIV:1101.0538;%%.

\bibitem{CMS-PAS-LUM-13-001}
\href {http://cds.cern.ch/record/1598864}{{{CMS}} Collaboration, ``{CMS
  Luminosity Based on Pixel Cluster Counting - Summer 2013 Update}'',} CMS
  Physics Analysis Summary CMS-PAS-LUM-13-001, 2013.

\bibitem{ATLAS:2011tau}
\href {https://cds.cern.ch/record/1375842}{{ATLAS and CMS} Collaboration,
  ``{Procedure for the LHC Higgs boson search combination in summer 2011}'',}
  Technical Report ATL-PHYS-PUB-2011-011, ATL-COM-PHYS-2011-818,
  CMS-NOTE-2011-005, 2011.

\bibitem{CLs}
\hrefCMSnoop {}{G.~Cowan, K.~Cranmer, E.~Gross, and O.~Vitells, ``Asymptotic
  formulae for likelihood-based tests of new physics'',} \textit{ Eur. Phys. J.
  C} \textbf{ 71} (2011) 1554,
  \href{http://dx.doi.org/10.1140/epjc/s10052-011-1554-0}{\doi{10.1140/epjc/s10052-011-1554-0}},
  \href{http://www.arXiv.org/abs/1007.1727}{\texttt{arXiv:1007.1727}}.

\bibitem{Denner:2011mq}
A.~Denner\hrefCMSnoop {}{ {et~al.}, ``Standard model {H}iggs-boson branching
  ratios with uncertainties'',} \textit{ Eur. Phys. J. C} \textbf{ 71} (2011)
  1753,
  \href{http://dx.doi.org/10.1140/epjc/s10052-011-1753-8}{\doi{10.1140/epjc/s10052-011-1753-8}},
\href{http://www.arXiv.org/abs/1107.5909}{\texttt{arXiv:1107.5909}}.
%%CITATION = ARXIV:1107.5909;%%.

\end{thebibliography}\endgroup

\cleardoublepage \appendix\section{The CMS Collaboration \label{app:collab}}\begin{sloppypar}\hyphenpenalty=5000\widowpenalty=500\clubpenalty=5000\textbf{Yerevan Physics Institute,  Yerevan,  Armenia}\\*[0pt]
V.~Khachatryan, A.M.~Sirunyan, A.~Tumasyan
\vskip\cmsinstskip
\textbf{Institut f\"{u}r Hochenergiephysik der OeAW,  Wien,  Austria}\\*[0pt]
W.~Adam, E.~Asilar, T.~Bergauer, J.~Brandstetter, E.~Brondolin, M.~Dragicevic, J.~Er\"{o}, M.~Flechl, M.~Friedl, R.~Fr\"{u}hwirth\cmsAuthorMark{1}, V.M.~Ghete, C.~Hartl, N.~H\"{o}rmann, J.~Hrubec, M.~Jeitler\cmsAuthorMark{1}, V.~Kn\"{u}nz, A.~K\"{o}nig, M.~Krammer\cmsAuthorMark{1}, I.~Kr\"{a}tschmer, D.~Liko, T.~Matsushita, I.~Mikulec, D.~Rabady\cmsAuthorMark{2}, B.~Rahbaran, H.~Rohringer, J.~Schieck\cmsAuthorMark{1}, R.~Sch\"{o}fbeck, J.~Strauss, W.~Treberer-Treberspurg, W.~Waltenberger, C.-E.~Wulz\cmsAuthorMark{1}
% ...
\vskip\cmsinstskip
% ...
1:~~Also at Vienna University of Technology, Vienna, Austria\\
2:~~Also at CERN, European Organization for Nuclear Research, Geneva, Switzerland\\
% ...

\end{sloppypar}
\end{document}